%% file: main.tex
\documentclass[lettersize,journal]{IEEEtran}

\usepackage{cite}
\usepackage{amsmath,amssymb,amsfonts}
\usepackage[ruled,linesnumbered]{algorithm2e}
\usepackage{graphicx}
\usepackage{textcomp}
\usepackage{xcolor}
\usepackage{url}
\usepackage{amsthm}
\usepackage[caption=false,font=footnotesize,labelfont=rm,textfont=rm]{subfig}

\usepackage{multirow}
\usepackage{booktabs}
\usepackage{pifont}
\usepackage[font =footnotesize ]{caption}
\usepackage{float}
\usepackage{booktabs}
\usepackage{threeparttable}
\usepackage{footnote}
\usepackage{siunitx}
\usepackage[T1]{fontenc}
\usepackage{inconsolata}
\usepackage{color}

\newtheorem{theorem}{\bf Theorem}

\newcommand{\liyao}[1]{\textcolor{black}{#1}}
\newcommand{\lyr}[1]{\textcolor{black}{#1}}

\IEEEoverridecommandlockouts
\begin{document}

\title{Shuffling for Semantic Secrecy
\thanks{Liyao Xiang (xiangliyao08@sjtu.edu.cn) is the corresponding author.}

\thanks{
This work was presented in part at IEEE GLOBECOM 2023 \cite{globecom} (doi: 10.1109/GLOBECOM54140.2023.10436975).  This work was partially supported by NSF China (62272306, 62032020, 62136006), NSFC with Grant No. 12426306, the Basic Research Project No. HZQB-KCZYZ-2021067 of Hetao Shenzhen-HK S\&T Cooperation Zone, NordForsk Nordic University Cooperation on Edge Intelligence under Grant No. 168043, and the Aarhus Universitets Forskningsfond project number AUFF 39001. Authors would like to appreciate the Student Innovation Center of SJTU for providing GPUs.}

\thanks{ Fupei Chen and  Liyao Xiang are with John Hopcroft
 Center for Computer Science, Shanghai Jiao Tong University, Shanghai
 200240, China (e-mail: cfp2022@sjtu.edu.cn; xiangliyao08@sjtu.edu.cn).}

 \thanks{Haoxiang Sun is with the School of Information and Electronic Engineering, Shanghai Jiao Tong University, Shanghai
 200240, China (e-mail: sunny\underline{~}sjtu@sjtu.edu.cn).}

 \thanks{Hei Victor Cheng is with the Electrical and Computer Engineering Department, Aarhus University, Denmark (e-mail: hvc@ece.au.dk).}

 \thanks{Kaiming Shen is with the School of Science and Engineering, The Chinese University of Hong Kong (Shenzhen), China (shenkaiming@cuhk.edu.cn).}

\thanks{This paper has supplementary downloadable material available at http://ieeexplore.ieee.org., provided by the author. The materials include mathematical proof processes, key-sharing methods, and additional experimental results. Contact cfp2022$@$sjtu.edu.cn for further questions about this work.}
}

% \thanks{A part of the work was presented in \cite{globecom}, published at the IEEE GLOBECOM 2023. This work was partially supported by NSF China (62272306, 62032020, 62136006), NSFC with Grant No. 62293482, the Basic Research Project No. HZQB-KCZYZ-2021067 of Hetao Shenzhen-HK S\&T Cooperation Zone, and the Aarhus Universitets Forskningsfond project number AUFF 39001. Authors would like to appreciate the Student Innovation Center of SJTU for providing GPUs.}
\author{\IEEEauthorblockN{
		Fupei Chen,
		Liyao Xiang, Haoxiang Sun, \\
  Hei Victor Cheng, and Kaiming Shen, \IEEEmembership{Senior Member, IEEE}}
          }
% \author{\IEEEauthorblockN{
% 		Fupei Chen\IEEEauthorrefmark{1},
% 		Liyao Xiang\IEEEauthorrefmark{1},
%         Hei Victor Cheng\IEEEauthorrefmark{2}, and Kaiming Shen\IEEEauthorrefmark{3}, Haoxiang Sun\IEEEauthorrefmark{1}}
% \IEEEauthorblockA{\IEEEauthorrefmark{1}John Hopcroft Center for Computer Science, Shanghai Jiao Tong University, Shanghai, China}
% \IEEEauthorblockA{\IEEEauthorrefmark{2}Electrical and Computer Engineering Department, Aarhus University, Denmark}
% \IEEEauthorblockA{\IEEEauthorrefmark{3}School of Science and Engineering, The Chinese University of Hong Kong (Shenzhen), China} \\
% \IEEEauthorblockA{E-mail: cfp2022@sjtu.edu.cn;  xiangliyao08@sjtu.edu.cn; hvc@ece.au.dk; shenkaiming@cuhk.edu.cn; sunny\underline{~}sjtu@sjtu.edu.cn}
% %\vspace{-3em}
% }

\maketitle

\begin{abstract}
Deep learning draws heavily on the latest progress in semantic communications. The present paper aims to examine the security aspect of this cutting-edge technique from a novel shuffling perspective. Our goal is to improve upon the conventional secure coding scheme to strike a desirable tradeoff between transmission rate and leakage rate. To be more specific, for a wiretap channel, we seek to maximize the transmission rate while minimizing the semantic error probability under the given leakage rate constraint. Toward this end, we devise a novel semantic security communication system wherein the random shuffling pattern plays the role of the shared secret key. Intuitively, the permutation of feature sequences via shuffling would distort the semantic essence of the target data to a sufficient extent so that eavesdroppers cannot access it anymore. The proposed random shuffling method also exhibits its flexibility in working for the existing semantic communication system as a plugin. Simulations demonstrate the significant advantage of the proposed method over the benchmark in boosting secure transmission, especially when channels are prone to strong noise and unpredictable fading. 
\end{abstract}


\input{introduction}
\input{relatedwork}

\input{architecture}

\input{background}

\input{security_analysis}
\input{experiment}
\input{conclusion}
\bibliographystyle{IEEEtran}
\bibliography{IEEEabrv,reference}

\begin{IEEEbiography}[{\includegraphics[width=1in,height=1.25in,clip,keepaspectratio]{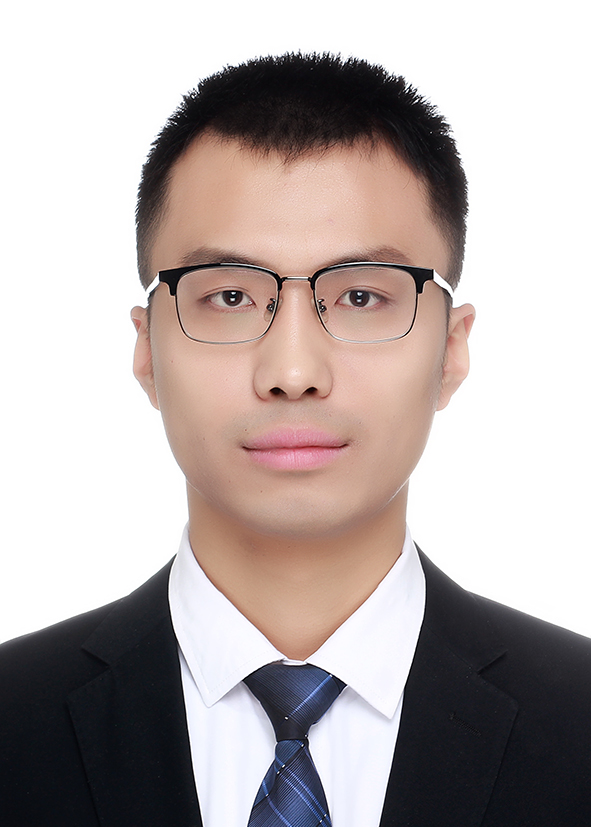}}]{Fupei Chen} received the B.S. degree in information science and engineering from Shandong University, Shandong, China, in 2022. He is working toward the Ph.D. in information and communication engineering at Shanghai Jiao Tong University. His research interests include security and privacy in semantic communication and learned image compression.
\end{IEEEbiography}

\begin{IEEEbiography}[{\includegraphics[width=1in,height=1.25in,clip,keepaspectratio]{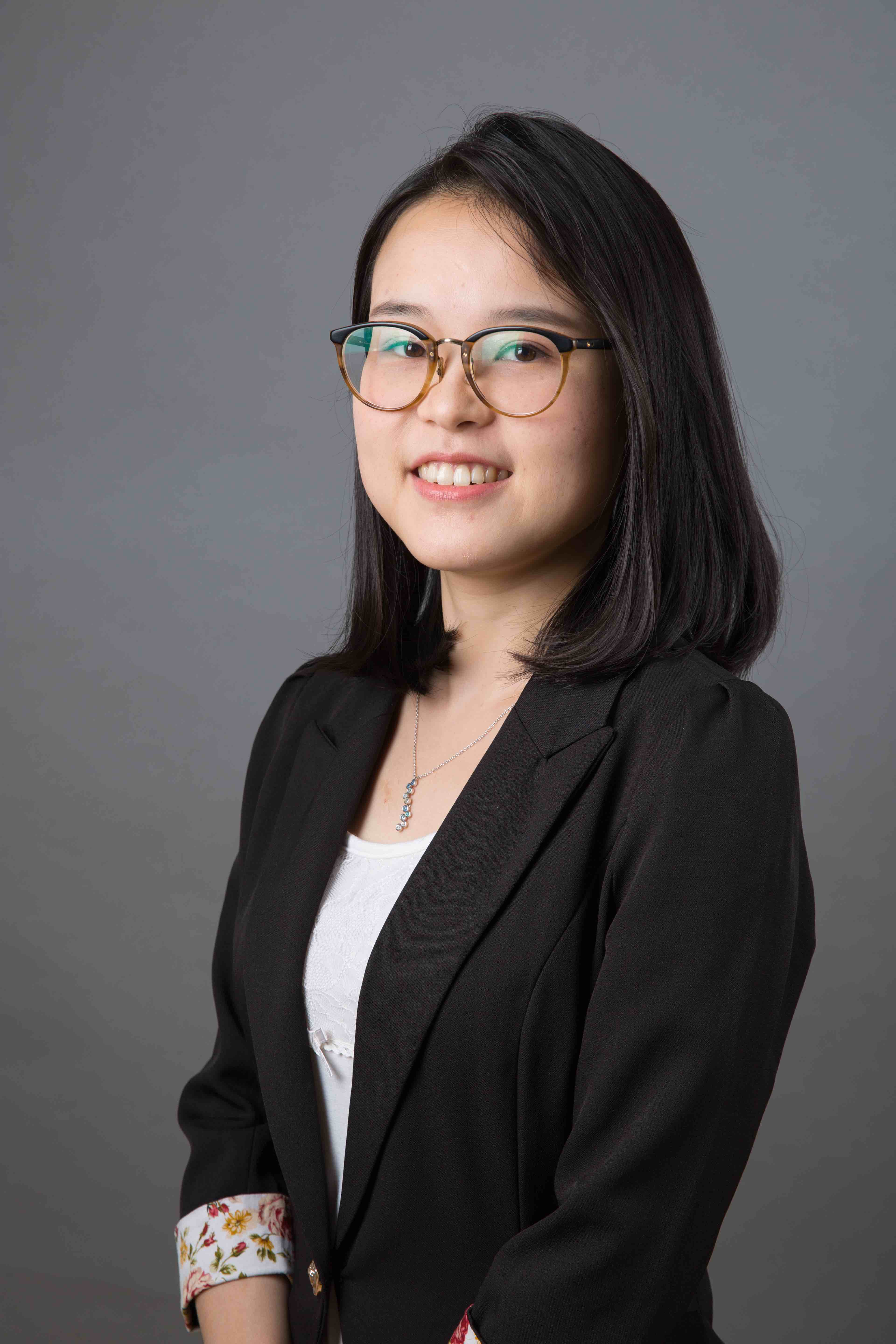}}]{Liyao Xiang} (Member, IEEE) received the B.Eng. degree in Electrical and Computer Engineering from Shanghai Jiao Tong University, Shanghai, China, in 2012, and the Ph.D. degree in Computer Engineering from the University of Toronto, Toronto, ON, Canada, in 2018. She is an associate professor at Shanghai Jiao Tong University. Her research interests include security and privacy, privacy analysis in data mining, and mobile computing.
\end{IEEEbiography}

\begin{IEEEbiography}[{\includegraphics[width=1in,height=1.25in,clip,keepaspectratio]{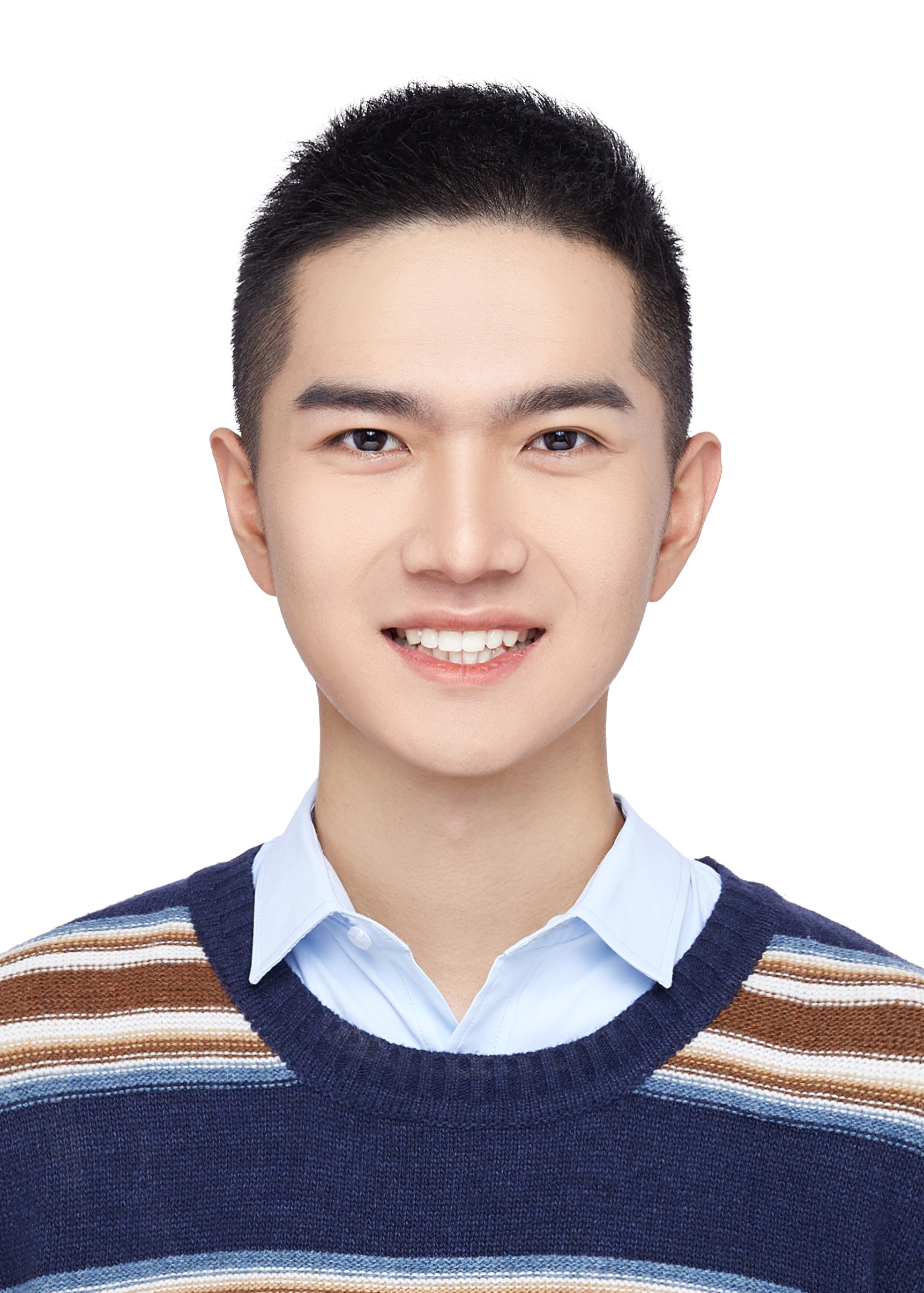}}]{Haoxiang Sun} is currently pursuing his B.S. degree in Computer Science at Shanghai Jiao Tong University, China. He will continue to work for his Master’s degree at the same institution. His research interests include semantic communication, natural language processing and large language models.
\end{IEEEbiography}

\begin{IEEEbiography}[{\includegraphics[width=1in,height=1.25in,clip,keepaspectratio]{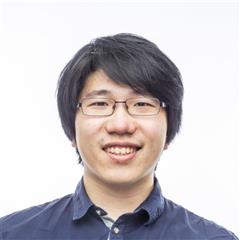}}]{Hei Victor Cheng}(Member, IEEE) received the B.Eng. degree in electronic engineering from Tsinghua University, Beijing, China, the M.Phil. degree in electronic and computer engineering from the Hong Kong University of Science and Technology, and the Ph.D. degree from the Department of Electrical Engineering, Linkoping University, Sweden. He worked as a Post-Doctoral Research Fellow at the University of Toronto, Toronto, ON, Canada. He is now an Assistant Professor with the Department of Electrical and Computer Engineering, Aarhus University, Denmark. His current research interests include next generation wireless technologies, intelligent surfaces, and machine learning.
\end{IEEEbiography}

\begin{IEEEbiography}
[{\includegraphics[width=1in,height=1.25in,clip,keepaspectratio]{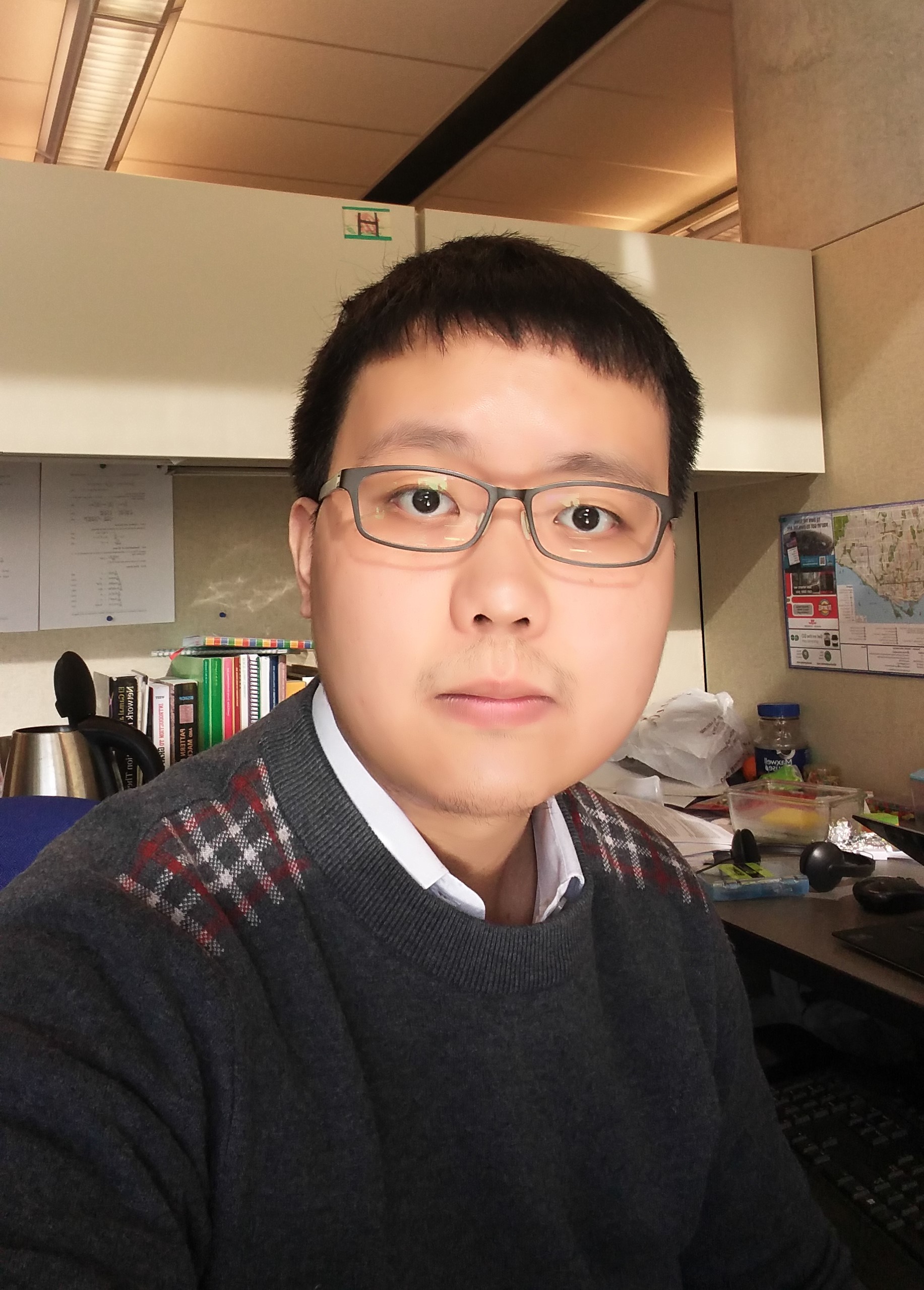}}]{Kaiming Shen}(Senior Member, IEEE) received the B.Eng. degree in information security and the B.Sc. degree in mathematics from Shanghai Jiao Tong University, China in 2011, and then the M.A.Sc. degree in electrical and computer engineering from the University of Toronto, Canada in 2013. After working at a tech startup in Ottawa for one year, he returned to the University of Toronto and received the Ph.D. degree in electrical and computer engineering in early 2020. Dr. Shen has been with the School of Science and Engineering at The Chinese University of Hong Kong (CUHK), Shenzhen, China as a tenure-track assistant professor since 2020. His research interests include optimization, wireless communications, information theory, and machine learning.
Dr. Shen received the IEEE Signal Processing Society Young Author Best Paper Award in 2021, the CUHK Teaching Achievement Award in 2023, and the Frontiers of Science Award at the International Congress of Basic Science in 2024. Dr. Shen currently serves as an Editor for IEEE Transactions on Wireless Communications.
\end{IEEEbiography}

\newpage
\clearpage
\input{supplement}
% \newpage
% \clearpage
% \input{review1}
% \input{review2}
% \input{review3}

\end{document}

%% file: introduction.tex
\vspace{-0.2cm}
\section{Introduction}

\begin{figure}[t!]
	% \vspace{-0.4cm}
	\centering
	\subfloat[traditional bit-based communication system]{
		\includegraphics[width=\columnwidth]{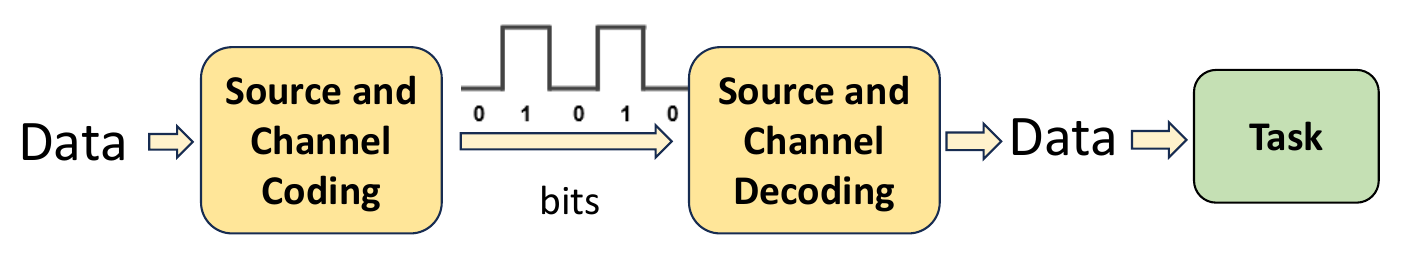}\label{awgn-simili}}
        \hfill
	\subfloat[semantic feature-based communication system]{
		\includegraphics[width=\columnwidth]{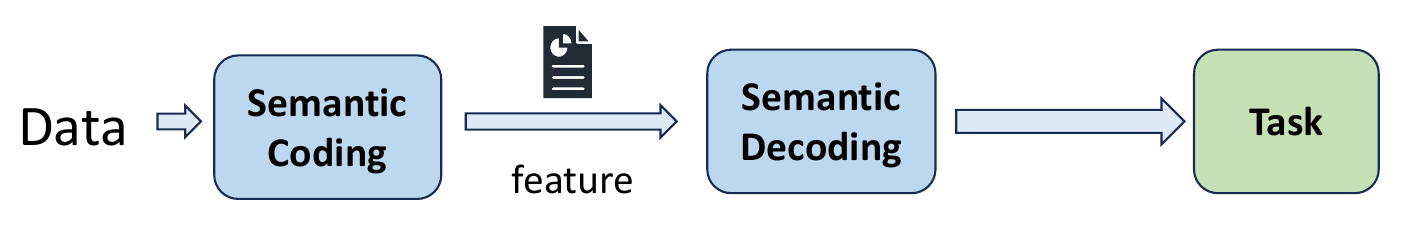}\label{ray-simili}}
	% \vspace{-0.3cm}
	\caption{Traditional communication system versus semantic communication system.}
	\label{comprasion image}
	\vspace{-0.4cm}
\end{figure}

% 1. First part: wiretap channel and theory secure:
Under the guidance of Shannon's seminal paper \cite{shannon} that laid the foundations of information theory, the primary pursuit of the modern communication system design has been efficient and accurate transmission of information bits. Artificial intelligence (AI) technology has found extensive application in autonomous driving, virtual reality (VR), augmented reality (AR), and related domains. 
However, the demands of these applications for transmission latency and precision surpass the capabilities of conventional communication methods. Consequently, there is a pressing need to develop communication solutions capable of handling these AI-driven systems. Concurrently, semantic communication \cite{semanticcommunication} has ventured into a new frontier in the area, which concentrates on the semantic essence and intention behind the information bits and thereby has the potential to improve transmission reliability and alleviate spectrum scarcity. Their difference is shown in Fig. \ref{comprasion image} wherein features, instead of bits, are transmitted, for data reconstruction or downstream tasks at the receiver end. 

In addition to transmission efficiency, the security of semantic information is under consideration. The broadcast trait of wireless channels renders wireless communications vulnerable to eavesdropping attacks, \liyao{where the attacker steals the private message of the sender to a legitimate party.} Despite semantic communication as a new communication paradigm, the semantic representation is still conveyed over the wireless channel, which is at high risk of eavesdropping attacks.

\liyao{Traditional security methods include encryption techniques \cite{RSA} and physical layer security \cite{PLSeurity} technologies in conventional communication systems. The high complexity of encryption hinders the former from being deployed in systems with real-time requirements. The physical-layer security typically relies on highly inefficient key generation and stringent channel conditions, limiting their potential applicability. Recent research on semantic communication systems also brings up security issues. As countermeasures, many works \cite{Luo,semprotector,sesc_21} adopt neural networks as encryption and decryption modules, apply adversarial learning, or explore physical-layer features. However, due to the black-box nature of neural networks, these methods fail to provide security guarantees while lacking efficiency.}

\liyao{To address the aforementioned issue, we propose a secure semantic communication system based on random shuffling the semantic features.} Our work stems from the conclusion that the noisy permutation channel would have a rate 0 by the conventional definition of transmission rate \cite{Polyanskiy}. The noisy permutation channel refers to a communication model in which an $n$-letter input undergoes a concatenation of a uniform permutation of these letters and a discrete memoryless channel. Inspired by it, we propose a noisy permutation channel from the semantic perspective: we devise a semantic joint source-channel coding scheme that extracts the semantic essence of raw information by a deep neural network (DNN) based encoder, and permutes the features before feeding them to the transmission channel. Considering features as symbols going through the transmission channel, an almost zero transmission rate would prevent the eavesdropper from decoding the permuted features. Meanwhile, those features can be exactly restored to their original order at a legitimate receiver knowing the permutation order, and are subsequently decoded into the message being sent. \liyao{The permutation equivariance of the neural networks for the encoder and decoder permits more flexibility in our system, enabling adaptability to different architectures and transmitted data types.} Unlike the neural network-based method, the proposed shuffling scheme does not incur any additional overhead for learning.

% We apply random permutation to the semantic features to simulate the uniform permutation channel. The permutation can be plugged into different places: it can be adopted before, after the semantic encoder, and after the channel encoder. The flexibility in choosing different shuffling positions allows varied trade-offs between the transmission rate and secrecy against the eavesdropper. What is more intriguing is that the permutation equivariance property of the Transformer-based encoder ensures that shuffling is equivalent at these three positions and all can be decoded by a legitimate receiver.

To instantiate such a semantic transmission architecture, we train the DNN-based encoder and decoder end-to-end with mutual information and leakage rate-based loss function to learn the optimal coding on the data distribution. This formulation incorporates the objectives of secrecy capacity, key rate, and leakage rates. \liyao{ We also provide a theoretical analysis of the upper bound of the leakage rate. }

Highlights of our contribution are as follows: we propose a secure semantic communication system that incorporates shuffling operation as a shared key to ensure the secrecy of the transmitted message against the eavesdropper while optimizing the secrecy capacity for the legitimate receiver. We adopt end-to-end training to optimize each module of the semantic communication system. Experiments on data from multiple modalities including images. text. and audio speech, show that our proposed approach is superior to traditional and other semantic approaches in enhancing transmission performance while suppressing information leakage. 

\liyao{
The rest of the paper is organized as follows. Section \ref{sec:related} reviews the related work. Section \ref{sec:allmodel} introduces the semantic communication system, the eavesdropping threats, and detailed description of the shuffling operation.
Section \ref{sec:trainstage} presents the concepts of secrecy capacity and leakage rate, and then describes the training process of the secure semantic communication system. Section \ref{sec:securityanalysis} provides a theoretical analysis of the security of the proposed system. Section \ref{sec:experiment} presents the experimental results with Section \ref{sec:conslusion} concluding the paper. The notations used in this paper are summarized in Table \ref{tab:notation}.
}

\begin{table}[h]
    \caption{\liyao{Notation list.}}
\centering
\resizebox{ \columnwidth}{!}{
    \begin{tabular}{ll}
\toprule % 加粗顶部横线
 Notation& Definition \\ \midrule 
\multirow{2}{*}{$\boldsymbol{M}, \boldsymbol{\widehat{M}}, \boldsymbol{M^{'}}$}  & message sent, message received by Bob, \\ & message received by Eve\\ 
\multirow{2}{*}{$\boldsymbol{U}, \boldsymbol{\widehat{U}}, \boldsymbol{U^{'}}$}  & input of semantic encoder, \\ & semantic decoder output of Bob, Eve \\ 
\multirow{2}{*}{
$\boldsymbol{T}, \boldsymbol{\widehat{T}}, \boldsymbol{T^{'}}$ } & input of channel encoder, \\ & channel decoder output of Bob, Eve  \\ 
$\boldsymbol{X}, \boldsymbol{Y}, \boldsymbol{Z}$  & data sent by Alice, data received by Bob and Eve \\ 
$T_{x}$  & neural networks of $x$
\\ 
$K_R, K_C$  & row key and column key \\ 
$\boldsymbol{P_R}, \boldsymbol{P_C}$  & row and column permutation matrix \\ 
$N, L, V, C$  & dimension of intermediate feature \\ 
$\boldsymbol{W}, \boldsymbol{W_{(R)}}$  & Transformer weights, Transformer weights shuffled \\ 
$P_X $  & transmission power constraint \\ 
$C_S, R_L$  & secrecy capacity, information leakage rate \\
$R_K, L_R$ & key rate, reconstruction loss \\ 
$\gamma, \alpha, \beta$  & row shuffle position, weight factors of $ R_L$ and $C_S$ \\ 
$n,g$  & the times of channel use, grain of shuffling \\
$\mathcal{M}, \mathcal{K}$  & message space and key space \\

\bottomrule % 加粗底部横线
\end{tabular}
}
\label{tab:notation}
\end{table}

%% file: relatedwork.tex
\section{Related Works}
\label{sec:related}

\subsection{Secure Transmission}
In contrast to the fruitful studies on the theoretical aspect of secure transmission, the practical implementation and algorithm design are somewhat lagging. According to Shannon \cite{informationtheorysecure}, the one-time pad preserves secrecy perfectly but is of limited practical value as the key string needs to be as long as the information string. Wyner considers reliable transmission over a discrete, memoryless channel subjected to a wiretap at the receiver, and builds an encoder-decoder such that the wiretapper's level of confusion is as high as possible \cite{wyner}. Ardestanizadeh et al. \cite{ArdesstanEhsan} enhance Wyner's coding scheme by incorporating a fresh randomness feedback strategy, thus establishing the feedback capacity of the wiretap channel. More recent works, including \cite{polarcode} and \cite{LDPC}, suggest using Polar codes and LDPC for secure transmission purposes, respectively.

For wiretap channel coding, Harrison et al. \cite{wiretapcode1} employ a three-stage encoder/decoder technique to implement a viable secret encoding scheme. Choi et al. \cite{wiretapcode2} focus on the secure coding problem in multiple-input multiple-output (MIMO) wiretap channels, utilizing secure pre-coding techniques to optimize the beam and artificial noise covariance matrix simultaneously, thereby improving the secure transmission rate of the system. %For wiretap channels with shared keys, key generation consists of four steps: channel measurement, quantification, information reconciliation, and privacy amplification. Received signal strength \cite{RSSsharedkey}, signal phase \cite{Phasesharedkey}, and Channel state information \cite{CSIsharedkey} are widely used channel characteristics. Advanced channel coding \cite{sharedkeyPolar,sharedkeyLDPC} is also employed in the information reconciliation stage to correct inconsistent bits in the keys generated by the two communicating parties. However, all these methods still incur high computational complexities.

\subsection{Semantic Communication}
Extensive research efforts have been devoted to this new communication paradigm which transmits different types of messages by interpreting the semantics with neural networks. In the following, we will introduce these works categorized by the type of message transmitted.

%text
For text transmission, Xie et al.\cite{deepsc} propose a bidirectional long short-term memory (BiLSTM) based coding scheme for semantic transmission, whereas Farsad et al. \cite{BiLSTM} resort to the Transformer for robust semantic transmission of the text. The work of Xie et al. \cite{LDeepSC} proposes a lightweight semantic communication system supporting IoT devices by pruning the model and adjusting the number of bits assigned to each model weight. Apart from point-to-point communication, Ma et al. \cite{RelayText} propose a deep learning-based semantic relay communication system. Jiang et al. \cite{HARQText} apply HARQ to semantic communication systems for transmitting texts, both improving the effectiveness and reliability of the system.

%image
Semantic communication systems for images mostly have CNN \cite{ImageSC15,ImageAdaptive16,ImageSlicemodel17,ImageTao18} or Transformer \cite{ImageVIT19,ImageSwin20} as the backbone. To mitigate the loss in image recovery caused by the gap between the distributions of the training and testing data, Zhang et al. \cite{ImageTao18} propose a data-adaptive network based on CGAN. Huang et al. \cite{ImageSC15} introduce a semantic encoding method depending on reinforcement learning, which allocates quantization bits by the importance of each pixel, significantly improving transmission efficiency.

Regarding speech transmission, DeepSC-S \cite{DeepSC-S} performs nonlinear transform and conditional coding to the input to extract the semantic feature of the speech data. The system of Han et al. proposed \cite{Speech2} jointly executes text transcription and speech recovery tasks, allowing the receiver to choose either the text or the audio signal to receive. Due to the diversity of voice data, spectrograms are widely used as inputs for processing \cite{Speech3SPECTGORM}, significantly reducing the communication resources required for transmission while ensuring performance.
%video

Given the substantial communication resource demand of video content, researchers have developed a semantic communication system for video transmission \cite{Video1,Video2,Video3}. This system employs deep learning to extract semantic information in the video frames, ensuring the quality of the video transmitted with an improved compression rate. 

\liyao{Other research efforts have been devoted to the application of semantic communication in multi-user systems. Zhang et al. \cite{MA_zhangs} categorized the knowledge of different users into shared and private information, where the former uses the shared channel to transmit while the latter uses dedicated channels, effectively saving transmission bandwidth. DeepMA \cite{DeepMA} is a multi-user semantic communication framework that saves bandwidth by encoding different data into mutually orthogonal semantic symbol vectors and transmitting them in parallel. Similarly, Liang et al. \cite{Liang_MA} introduced the Orthogonal-Model Division Multiple Access (O-MDMA) technology, leveraging the non-interchangeable nature of semantic information across different semantic models to exploit the orthogonality of semantic information, further optimizing bandwidth utilization. }
% In addition, researchers have also attempted to apply Non-Orthogonal Multiple Access (NOMA) to semantic communication systems \cite{NOMA1, NOMA2} to improve the performance bound even further.
Nevertheless, the aforementioned works overlook the data security issue in semantic communication systems which is our target.

%In this work, we aim to fill this gap by proposing a novel semantic communication framework with shuffling order serving as the shared key. In experiments, our framework recovers the semantic meaning with high accuracy while preventing eavesdropper reconstruction. Our system outperforms traditional secrecy communication systems and improves robustness at the low SNR regime.

%1. semantic communication
\begin{figure*}[h!]
    \centering
    %\vspace{-0.3cm}
    \includegraphics[width=\linewidth]{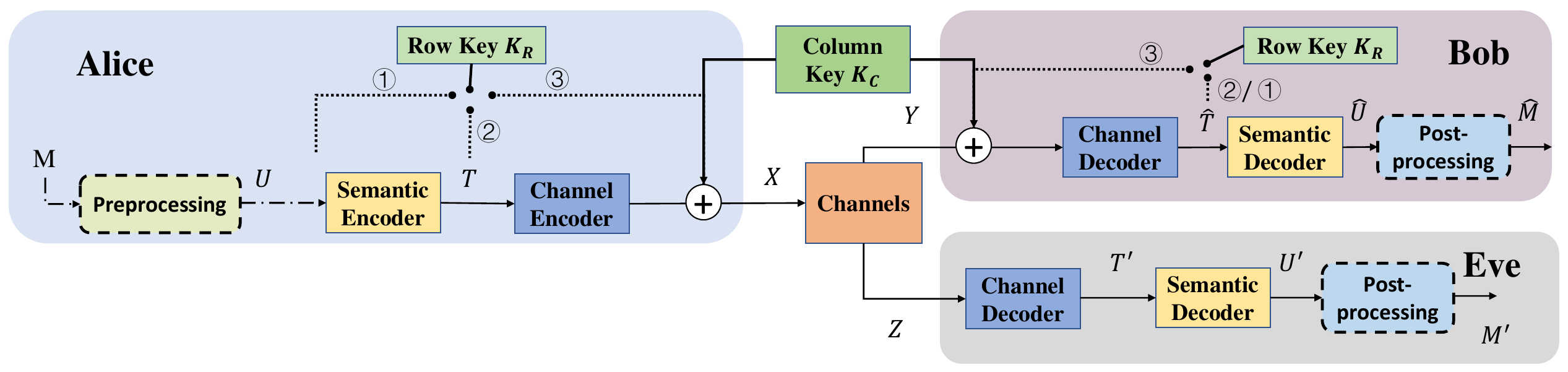}
    \caption{Secure communication over a public channel with a shared key. The symbol $K_R, K_C$ represent the row shuffling key and column shuffling key, respectively. The dotted line denotes the position where the row shuffling/inverse shuffling takes place. The preprocessing and postprocessing differ by different types of data transmitted.}
    \label{system image}
    \vspace{-0.2cm}
\end{figure*}

\subsection{Secure Semantic Communication}
There are comprehensive studies on security issues in semantic communication \cite{sesc_suvey1, sesc_survey2, sesc_survey3}, generally categorized into two aspects: the security of neural network models and the security of transmitted data. \lyr{The former examines the vulnerabilities of neural networks in communication. For instance, Nan et al. \cite{paper5} explored the impact of adversarial perturbations on semantic communication systems. By injecting adversarial perturbations into the data transmitted over the air, they effectively misled the downstream classification tasks at the receiver. The latter focuses on preventing unauthorized information interception by attackers.} This work focuses on the latter, particularly on preventing eavesdroppers from recovering transmitted data.

Among these works, most of them adopt neural networks as encryption and decryption modules. Luo et al. \cite{Luo} designed an adversarial encryption training scheme, utilizing a neural network to generate the key for data encryption. Liu et al. \cite{semprotector} also applied neural networks in encryption and decryption modules, which could serve as a plug-in enabling flexible integration in various semantic communication systems. Wang et al. \cite{sesc_21} utilized information bottleneck and adversarial learning to defend against eavesdropping attacks, where the former extracts task-relevant features from the input, and the latter prevents attackers from reconstructing the original data from these features. He et al. \cite{sesc_14} used adversarial perturbations to prevent reconstruction attacks while designing a decoder at the receiver to eliminate the impact of perturbation. Additionally, some methods \cite{sesc_nn, sesc_nn2, sesc_nn3} leveraged the inherent complexity of neural networks to protect the transmitted data, but due to the irreversibility and black-box attribute of neural networks, these methods not only compromise transmission efficiency but also require further validation of their security guarantees. In contrast, steganography techniques were employed by Tang et al. \cite{sesc_9} to hide confidential images in the transmitted messages. 

Several training-free schemes have also been proposed. Based on the public-key encryption scheme, Tung et al. \cite{DeepJSCEC} proposed a wireless image transmission system that provides security against chosen-plaintext attacks from the eavesdropper. Qin et al. \cite{encryptionandobfusion} proposed a novel physical-layer semantic encryption scheme by exploring the randomness of bilingual evaluation understudy (BLEU) scores in the field of machine translation and a subcarrier obfuscation mechanism to provide further physical layer protections. A superposition coding-based secure semantic communication system was introduced by \cite{sesc_18}, including a constellation map that prevents eavesdroppers from obtaining any valid information about private data. The system of \cite{sesc_20} simultaneously transmits semantic information and bit information, with the former as information-bearing artificial noise to interfere with the conventional bit-oriented communication channel of the attacker. \lyr{Rong et al. \cite{paper22} proposed a semantic entropy-guided encryption mechanism, which utilizes semantic scores to generate keys via a hash function, ensuring compatibility with various semantic communication systems.} However, the aforementioned methods either fail to provide sufficient security guarantees or are highly inefficient, limiting their practical application.

\liyao{Random permutation and substitution are proposed in \cite{sesc_1} as a defense as opposed to the model inversion eavesdropping attack, but they did not exploit any neural network property to ensure computation equivalence, nor analyze the channel from a noisy permutation channel perspective. Compared with their work, ours emphasizes the transmission capacity and leakage rate with optimizable losses.}

%% file: architecture.tex
\section{The Secure Semantic Communication Framework} 
\label{sec:allmodel}

\liyao{In this section, we first introduce our proposed secure semantic communication system and the related security threats. Then, we show the shuffling scheme in detail, and how it could be generalized to other communication systems.}

\subsection{A Secure Semantic Communication System}
\label{sec:system_model}
Fig.~\ref{system image} depicts the general point-to-point communication system with an eavesdropper under consideration. The sender Alice wishes to send a message to the receiver Bob while keeping it secret from the eavesdropper Eve. The semantic encoder learns to extract the meaning of the input, compress it, and `encrypt' it by shared key $K$, whereas the semantic decoder learns to reconstruct the input by `decrypting' the received stream.  Without the secret key, Eve deciphers the message by random guessing. The goal of the system is to minimize semantic errors while reducing the number of symbols to be transmitted between legitimate sender and receiver, yet without leaking any information to an eavesdropper. Different from existing secure communication systems for bit-level transmission, we propose joint source and channel coding which permits transmission/leakage at the semantic level, i.e., if the semantic recovery fails due to bit loss, the transmission is considered unsuccessful.
% Both channel encoder and decoder are learned from end to end and are adjusted according to the data distribution to best match the channel condition.

To achieve successful semantic recovery and prevent leakage, we jointly design the sender and receiver by DNNs with a shared key to enable legitimate transmission while suppressing eavesdropping. Specifically, we assume that the input of the communication system is text and we design the encoder to incorporate a semantic encoder concatenated by a channel encoder where the former is composed of Transformer encoder blocks $Enc$ and the latter consists of MLP layers. The secure semantic transmission system is depicted in Fig.~\ref{system image}.

% The sender maps the content to be sent $M={[w_{1},w_{2},...w_{\mathcal{M}}]}$ where $w_{i} \in \mathcal{M}$ denotes the $i$-th token sampled from the message space $\mathcal{M}$ into a symbol stream with a shared key $K$, and then passes it through the physical channel with impairments. The receiver decodes the received data with $K$ to reconstruct 
the original content. Meanwhile, the eavesdropper also attempts to decode the received to recover the sentence without $K$. 
The sender Alice pad the content to be transmitted to a fixed length $N$ to constitute the input $\boldsymbol{M}={[w_{1},w_{2},...w_{N}]}$ which is further converted into encoding $\boldsymbol{U}\in \mathbb{R}^{N\times L}$ through an embedding layer, where $L$ is the dimension of embedding vector and this step corresponds to the preprocessing module in Fig.~\ref{system image}. The input embedding is fed into a semantic encoder to output $ \boldsymbol{T}\in \mathbb{R}^{N\times L}$ which is sent to the channel encoder.  To ensure secure transmission, the message needs to be encrypted using the row key $K_R$ and the column key $K_C$ before transmission over the channel. Due to the unique properties of neural networks, the row key can be applied at three different positions on the sender's Alice, controlled by the parameter $\gamma$, while the column key can only be applied to the output of the channel encoder. We will provide a detailed explanation of this process in the next section. \liyao{To meet power constraints, the message $\boldsymbol{X} \in \mathbb{R}^{N\times V}$ will be normalized before transmission, where $V$ is the feature dimension. The whole encoder process can be expressed as:}
\liyao{
\begin{subequations}
    \begin{align}
    \label{eq:encoder}
    \boldsymbol{X} = T_{Alice}(\boldsymbol{M}, K_R, K_C, \gamma),\\ \textrm{s.t.} ~~\frac{1}{NV}\sum_{i = 1}^{NV} {||x_i||}^2 \leq P_X,
    \end{align}
\end{subequations}}

\liyao{where \(T_{Alice}\) is the encoder function, $ x_i$ is the $i$-th element of $\boldsymbol{X}$, and $P_X$ denotes the average transmitting power constraint. We set $P_X =1$ throughout the paper.}
Alice transmits $\boldsymbol{X}$ through the physical channel and Bob receives $\boldsymbol{Y}$, a noisy version of $\boldsymbol{X}$. The eavesdropper receives $\boldsymbol{Z}$ through a wiretap channel which we assume to have the same channel condition as Bob.  
Symmetric to the sender, the receiver Bob feeds $\boldsymbol{Y}$ through a channel decoder and then a semantic decoder while deciphering it at the corresponding positions with the shared key ${K_R}$ and ${K_C}$, and finally reconstruct the message $\boldsymbol{\hat{M}}$.  The decoding process of Bob can be represented as:
\liyao{
\begin{equation}
\label{eq:decoder}
\boldsymbol{\hat{M}} = T_{Bob}(\boldsymbol{Y}, K_R, K_C, \gamma),
\end{equation}
where \(T_{Bob}\) is the Bob's decoder function. Meanwhile, the eavesdropper Eve directly decodes the received data $\boldsymbol{Z}$ with its decoder to obtain $\boldsymbol{M'}$. The reconstruction process of Eve can be represented as:
\begin{equation}
\boldsymbol{M'} = T_{Eve}(\boldsymbol{Z}),
\end{equation}
where \(T_{Eve}\) is the Eve's decoder function.}

%Symmetric to the sender, the receiver (the eavesdropper) feeds $Y$($Z$) through a channel decoder and then a semantic decoder to reconstruct the input $\hat{M}$($M'$).
%B: threat model

\liyao{In the following we describe the security threats of the communication system.}

\textbf{Attack scenarios:} We consider inversion attacks, where eavesdropper Eve tries to recover the original message $\boldsymbol{M}$ from the received $\boldsymbol{Z}$. \lyr{Previous works \cite{inversionatt1,inversionatt2} have demonstrated that inversion attack by neural networks is an effective approach, where a neural network is trained to invert features. Therefore, in the semantic communication system illustrated in Fig.~\ref {system image}, the eavesdropper Eve trains a decoding network $T_{Eve}$ to reconstruct the message $\boldsymbol{M'}$ from the received features $\boldsymbol{Z}$.}
The goal of the attack is to minimize the reconstruction error between the original message $\boldsymbol{M}$ and the recovered message $\boldsymbol{M'}$.

\liyao{
\textbf{Attacker's prior:} We consider a powerful attacker Eve who can access the training datasets of the legitimate encoders and decoders and can eavesdrop on all the messages on the transmission channel. Consequently, Eve can train a decoder $T_{Eve}$ to recover the original input of Alice's. We assume that the decoder network structure and the channel condition are the same as Bob's. The only disadvantage of Eve compared to Bob is that it does not know the secret key. Hence, Eve relies on training its decoder to crack the message in transmission.}

\begin{figure*}[ht!]
    \centering
    \subfloat[]{
        \includegraphics[height=4.5cm, keepaspectratio]{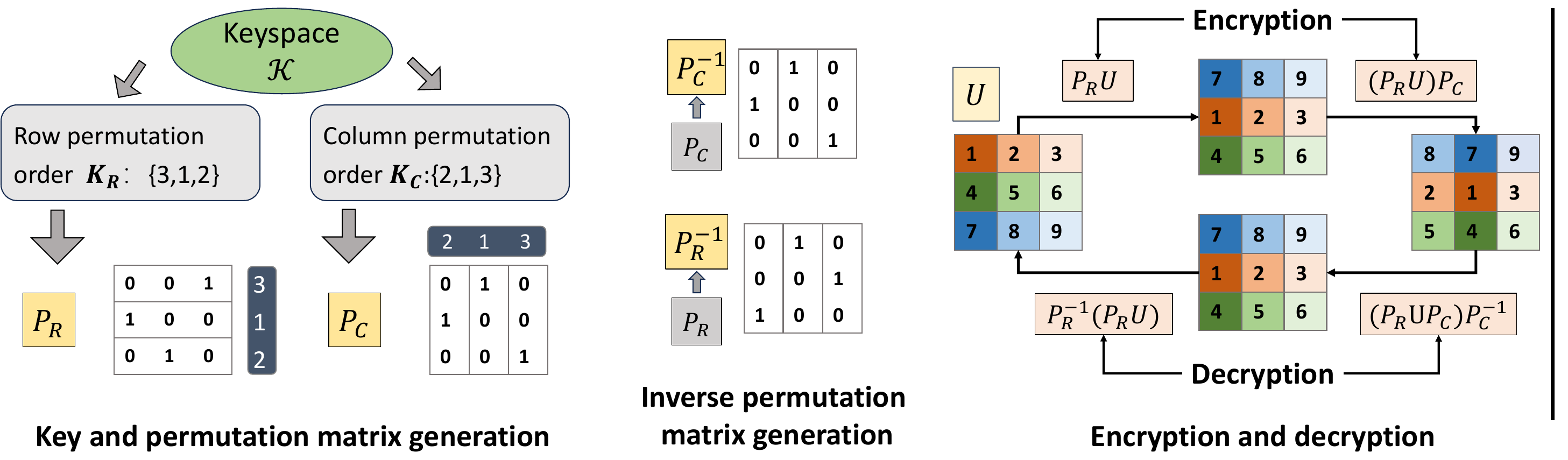}\label{shuffleoperation}}
    \hfill
    % \centering
    \subfloat[]{
        \includegraphics[height=4.5cm, keepaspectratio]{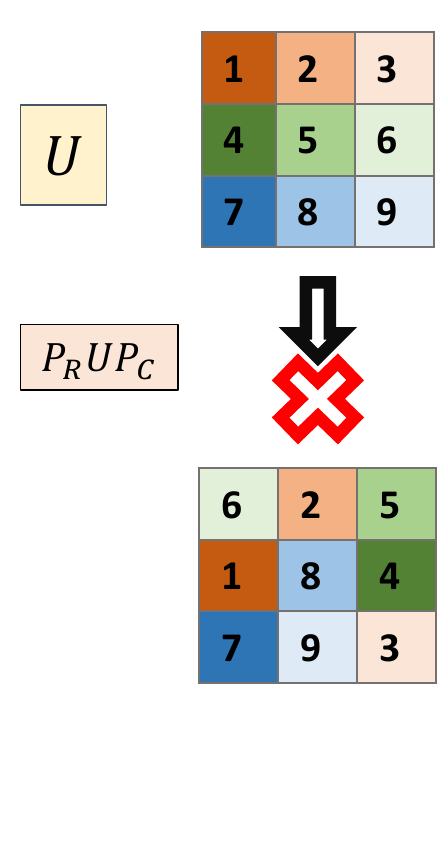}\label{denyshuffle}}
    \caption{\liyao{(a) The detailed illustration of permutation and inverse permutation. (b) An example of a single row permutation and column permutation cannot obtain an arbitrary arrangement pattern. }}
    \label{opeartion} 
    \vspace{-0.4cm}
\end{figure*}

\subsection{The Permutation Key}
\textbf{Motivation:} The key shared between the sender and the legitimate receiver is essential to secure communication. We design the key to be the \textit{permutation order} of the intermediate representation of the encoder output, inspired by the noisy permutation channels \cite{Polyanskiy}. Noisy permutation channels are introduced in \cite{NoisyPermutationChannel} in which an $n$-symbol input undergoes a concatenation of a discrete memoryless channel and a uniform permutation of the $n$ letters, motivated by the application of the multipath routed network and DNA storage system. Since the receiver observes a uniformly permuted output, the order of symbols conveys no information. Under the common definition of rate $R = \frac{\log \mathcal{|M|}}{n}$, the noisy permutation channel would have rate $0$ by \cite{Polyanskiy}. Unless stated otherwise, we use $\log(\cdot)$ to denote logarithm base 2. In the semantic communication setting, the message in the transmission is the intermediate representation, typically of high dimension, providing ample space for feature permutation. \liyao{Additionally, the structural information in images and the contextual relationships in text are closely related to the order of intermediate features. Disrupting this order can destroy the dependency relationships and consistency in data, leading to the failure of decoding. For example, a word sequence of different orders suggests dramatically different semantic meanings.} Thus, we randomly permute the input encoding $\boldsymbol{U} \in \mathbb{R}^{N \times L}$, which resembles a noisy permutation channel to the unauthorized party Eve so that Eve has an almost $0$ rate.

\liyao{
The key $K=(K_R, K_C)$ includes two independent keys $K_R$ for row permutation and $K_C$ for column permutation. The keys are randomly sampled from all possible permutations. For $\boldsymbol{U}\in \mathbb{R}^{3\times 3}$, a total of $3! = 6$ permutations are possible for row shuffling. We then derive row permutation matrix $\boldsymbol{P_R}$ by permuting the row of an identity matrix according to $K_R$. Column permutation matrix is obtained accordingly. Since the intermediate features of neural networks are mostly high-dimensional tensors, we choose matrix multiplication for shuffling, as shown by the example of $\boldsymbol{U}\in \mathbb{R}^{3\times 3}$ in Fig.~\ref{shuffleoperation}. Row shuffling for $\boldsymbol{U}$ is done by $\boldsymbol{P_R} \boldsymbol{U}$ and column shuffling is done by $\boldsymbol{U} \boldsymbol{P_C}$. Notably, the two shuffling methods have different indications that row permutation means rearranging inter-token positions whereas column permutation suggests intra-token shuffling. Correspondingly, a legitimate receiver multiplies the inverse of permutation matrices $\boldsymbol{P_R^{-1}}$ and $\boldsymbol{P_C^{-1}}$ to recover $\boldsymbol{U}$.}

It is worth pointing out that the combination of row and column permutation merely covers a subspace of the entire permutation space. An exclusion is provided in Fig.~\ref{denyshuffle}. Hence for a $3 \times 3$ matrix, there are in total $3! \times 3!$ possible permutations instead of $9!$ element-wise permutations. This would mildly affect the key space as discussed below.
% We permute the input embeddings by multiplying a permutation matrix $P_R \in \left \{ 0,1 \right \}^{N \times N }$ where $P_R U$ stands for a row-shuffled $U$. Correspondingly, column shuffling is denoted by $U P_C$ where $P_C \in \left \{ 0,1 \right \}^{L\times L}$. 

\textbf{Permutation position:}
According to \cite{hengyuanPermute}, both the Transformer encoder and the MLP layer have row permutation forward equivalence and backward invariance: 
\liyao{
    \begin{subequations}\label{xhy}
    \begin{align} 
        Enc(\boldsymbol{P_{R}U}) &= \boldsymbol{P_{R}}Enc(\boldsymbol{U}),\\
       \frac{ \partial L}{\partial \boldsymbol{W}} &= \frac{ \partial L}{\partial \boldsymbol{W_{(R)}}},
    \end{align}    
    \end{subequations}}
where $Enc$ is the Transformer encoder, $L$ is loss function, and $\boldsymbol{W}$ and $\boldsymbol{W_{(R)}}$ are the weights of Transformer when taking $\boldsymbol{U}$ and $\boldsymbol{P_{R}U}$ as input, respectively. The results demonstrate that the output or the gradient of the Transformer encoder under shuffling would remain the same with the counterparts without shuffling. The shuffling invariance property, as pointed out by \cite{hengyuanPermute}, extends beyond the Transformer architecture to a wide range of neural networks.

Considering the impact of shuffle operation on the input reconstruction, we restrict the application of column permutation to the output of the channel encoder only, and apply row shuffling to one of the following: 1) output of the embedding layer ($\boldsymbol{U}$); 2) output of the semantic encoder ($\boldsymbol{T}$); and 3) output of the channel encoder ($\boldsymbol{X}$), as shown in Fig.~\ref{system image}. Due to the shuffling invariance property, the choice of the row shuffling position has a minor impact on the system performance. 

% Hence a noisy permutation channel is created for the eavesdropper. By \cite{Polyanskiy}, without the shared key, the rate for the transmission channel of Eve is almost zero and hence Eve can hardly reconstruct the original message.

In particular, due to the permutation equivariance, the row permutation is equivalent at positions 1, 2, and 3 in Fig.~\ref{system image} and all could be removed by inverse shuffling.

\liyao{
\textbf{Shuffle Grain:} We define grain as the smallest unit of permutation, denoted by $g$. For example, consider indices set $\{1,2,3,4,5,6\}$, $g = 1$  means that each element can be moved to any other position, resulting in $6!$ possible permutations. If $g = 2 $, every two elements are treated as a unit, i.e.,  $\{1, 2\}, \{3, 4\}, \{5, 6\}$, resulting in $3!$ possible permutations. Typically, $g$ is a factor of the size of the indices set. Unless otherwise specified, we set $g = 1$ by default.}

\liyao{\textbf{Key Sharing:}}
We assume that both Alice and Bob own the same codebook, and the codebook consists of a set of \emph{`index-key'} pairs, where \emph{key} denotes the permutation order, and each index corresponds to each key. The codebook only contains the encryption key since the decryption key could be easily obtained from the encryption key by taking the inverse. Note that the specific form of the key depends on the shape of feature $\boldsymbol{X}$ to be encrypted, which relies on the encoder structure. Before the transmission of the actual content, legitimate parties use the Diffie-Hellman key exchange protocol to generate a random index within the codebook. With the index, Alice and Bob reached an agreement on the key that they would use in the following transmission. \liyao{Please refer to Supplementary Materials-C for the detailed description.}
\liyao{\subsection{Universality of the Shuffling Scheme}
Since the shuffling operation is applied to the data to be transmitted and is independent of the data type and communication mode, it is suitable for a wide variety of modalities and communication settings.
}

\textbf{Multi-Modal Semantic Communication:} Our secure semantic communication framework with shuffle could serve data of different modalities. Hence we instantiate our framework with three types of data, i.e., text, image, and audio stream, respectively. For different types of data, the specific meaning of each notation and the model structure in the framework of Fig.~\ref{system image} varies. For text transmission, the token $w_i$ represents the $i$-th word while it denotes the $i$-th patch in image transmission and the $i$-th sample point in the speech audio transmission.

For both text and image transmission, we adopt Transformers as the backbone of the network. \liyao{Particularly, we adopt ViT \cite{Vit} in the image transmission, where the image} $M$ is divided into patches that are embedded whereas the text sentence is embedded at the word level; likewise, the patches are decoded at the receiver are required to be reinstalled to reconstruct the image. \liyao{The division and reinstallation operations are described as pre-processing and post-processing modules in Fig.~\ref{system image}.}
% The image transmission differs from the text transmission in that each input image $M$ is divided into patches that are embedded whereas the text sentence is embedded at the word level; likewise, the patches decoded at the receiver are required to be reinstalled to reconstruct the image.

For speech transmission, both semantic encoding/decoding and channel encoding/decoding are composed of convolutional neural networks (CNN), which differ from Transformer-based ones. First, the row shuffling can only be applied to the output of the channel encoder as the convolution operation fails to satisfy the permutation equivariance \cite{hengyuanPermute}. Second, audio data requires preprocessing which frames the audio speech before being fed into the semantic encoder. At the receiver, the decoding includes a de-frame operation to recover the original signal. \liyao{Notably, since the audio feature $\boldsymbol{X} \in \mathbb{R}^{C\times N\times L}$ ($C$ is the number of filters and $N, L$ is the height and width of the feature, respectively) in transmission is of higher order than image or text, it not only could use row and column shuffling, but also be permuted at the filter level.} 

\liyao{
\textbf{Other communication scenarios:} Apart from point-to-point wireless communication, our approach is widely applicable to a wide range of scenes, such as MIMO, cellular networks, and wired communication. For instance, in the uplink, each user can encrypt their data with their unique key and then send it to the base station. The base station receives data from users and decrypts messages using keys. Similarly, in the downlink, the base station can distribute private data to different users using different keys, ensuring each user only decrypts their share of data. Even in wired communication scenarios, such as internal communication within data centers, the random shuffling operation can provide an additional layer of security.}

%% file: background.tex
\section{Secrecy Capacity}
\label{sec:trainstage}

\liyao{Having illustrated our framework, we now show how each module is trained. Prior to that, we introduce the concept of secrecy capacity from information theory, leveraging it to analyze the security of our proposed system.}

\subsection{Secrecy Capacity and Key Rate}
\textbf{Secrecy Capacity:} To describe the goal of secure communication between Alice and Bob in Fig. \ref{system image}, we use the secrecy capacity \cite{NIF}, referring to the maximum transmission rate achieved by legitimate parties with a shared key $K$, while ensuring near-zero information leakage rate $R_L$:
\begin{subequations}\label{CS1}
\begin{align} 
      C_{S}(R_{K}) &= \max_{p(x)} \left\{I(\boldsymbol{X};\boldsymbol{Y})-[I(\boldsymbol{X}; \boldsymbol{Z})-R_{K}]^+\right\},\\
    &\quad\;\;\textrm{s.t.}\lim \sup_{n \rightarrow \infty} R_L=\frac{1}{n}I(\boldsymbol{M};\boldsymbol{Z}^{n}) \leq \epsilon,
\end{align}    
\end{subequations}
where $I(\cdot;\cdot)$ is the mutual information, $\epsilon$ is an arbitrarily small positive number, and $[x]^+=\max\{x,0\}$. $\boldsymbol{X}$, $\boldsymbol{Y}$, and $\boldsymbol{Z}$ denote the message random variables sent by Alice, received by Bob, and received by Eve, respectively. $R_K$ denotes the transmission rate of the key. $\boldsymbol{Z}^{n}$ represents $n$ times channel use (by which a symbol is transmitted per use) and $n$ is the length of the symbol stream.

The achievability suggests that when the transmission rate $R<C_S$, there exists a coding scheme that allows the legitimate receiver to recover the original input with arbitrarily small error probability while preventing the information from being leaked to the eavesdropper. Thus, the security of the communication system is measured by the information leakage rate ($R_L$), while the secrecy capacity $C_S$ quantifies reliability.

In practice, it is hard to evaluate the mutual information as in Eq.~\eqref{CS1} directly, so we resort to a neural estimator \cite{mine2018} to obtain an approximate evaluation. Specifically, mutual information can be rewritten as the Kullback-Leibler (KL) divergence between the joint probability density ($\mathbb{P}_{\boldsymbol{X Y}}$) and the product of the marginal probabilities ($\mathbb{P}_{\boldsymbol{X}} \mathbb{P}_{\boldsymbol{Y}}$):
\begin{equation}
    I(\boldsymbol{X} ; \boldsymbol{Y})=D_{\mathrm{KL}}(\mathbb{P}_{\boldsymbol{X Y}}\|\mathbb{P}_{\boldsymbol{X}} \mathbb{P}_{\boldsymbol{Y}}),  %\otimes 
\end{equation}
Based on the Donsker-Varadhan variational representation, the KL divergence has the following dual representation: 
\begin{equation}
    D_{K L}(\mathbb{P}_{\boldsymbol{X Y}} \| \mathbb{P}_{\boldsymbol{X}} \mathbb{P}_{\boldsymbol{Y}})=\sup _{F: \Omega \rightarrow \mathbb{R}} \mathbb{E}_{\mathbb{P}_{\boldsymbol{X Y}}}[F]-\log \left(\mathbb{E}_{\mathbb{P}_{\boldsymbol{X}} \mathbb{P}_{\boldsymbol{Y}}} \left[e^{F}\right]\right),   %\Omega
\end{equation}
where the supremum is taken over all functions $F: \Omega = \mathbb{X} \times \mathbb{Y} \rightarrow \mathbb{R}$ such that the two expectations are finite. For any given function $F$, the right-hand side of the aforementioned expression corresponds to a lower bound on mutual information $I(\boldsymbol{X}; \boldsymbol{Y})$.

Similar to \cite{deepsc}, a neural network $T_{\theta}$ parameterized by $\theta$ is used in substitution of $F$. Hence we can maximize the lower bound to estimate mutual information:
\begin{equation}\label{eqmine}
   L_{MI}^{\theta}(\boldsymbol{X};\boldsymbol{Y})=\mathbb{E}_{\mathbb{P}_{\boldsymbol{X Y}}}\left[T_{\theta}\right]-\log \left(\mathbb{E}_{\mathbb{P}_{\boldsymbol{X}} \mathbb{P}_{\boldsymbol{Y}}}\left[e^{T_{\theta}}\right]\right).
\end{equation}

% \begin{align}
% 	I(X ; Y) \geqslant L_{MI}^{\theta}(X;Y)=\mathbb{E}_{\mathbb{P}_{X Y}}\left[T_{\theta}\right]-\log \left(\mathbb{E}_{\mathbb{P}_{X} \mathbb{P}_{Y}}\left[e^{T_{\theta}}\right]\right).
	
% \end{align}

%%%%%%

\liyao{
\textbf{Key Rate $R_K$:} Aside from the mutual information, the key rate $R_K$ in Eq.~\eqref{CS1} also needs to be calculated. 
The permutation order acts as the shared key $K=(K_R, K_C)$, and keyspace $\mathcal{K}$ contains all possible permutation indices. For the semantic feature $\boldsymbol{X}\in \mathbb{R}^{N\times L}$, there are $N!$ possible row permutations and $L!$ possible column permutations, making up a space of size $| \mathcal{K} | = N!L!$. For the semantic feature $\boldsymbol{X}\in \mathbb{R}^{C \times N\times L}$ in the CNN-based network, there are $N!$, $L!$, $C!$ likely permutations for row shuffling, column shuffling, and filter shuffling, respectively, comprising a total space of size $| \mathcal{K} | = N! L! C!$. }

Note that the conventional unit for $R_K$ is bits per symbol and the key length required for transmission is considered. In the realm of semantic communication, the basic unit of transmission is the smallest meaningful unit to be transmitted. \liyao{For text data, a basic unit is a word, for image data, it is a patch, and for audio data, it is a sampling point. Thereby we have:}

% \begin{definition}\label{def:keyrate}
% For $n$ symbols transmitted and their input embedding $U$ permuted by $K \in \mathcal{K}$, the key rate is given by%$R_{K}$ is
% \begin{equation}
%     R_{K} =\frac{\log (|\mathcal{K}|)}{n}.
% \end{equation}
% \end{definition}

% \subsection{The Key Space}
% Since the key space differs by the type of encoder-decoder network structure, the key rate by Def.~\ref{def:keyrate} is different. Note that $R_K$ is a part of Eq.~\eqref{CS1}, which has the same unit as the mutual information $I(X; Y)$. Such a unit is defined by the granularity of the symbol transmitted in semantic communication. Specifically, we have
\begin{theorem}
For transmitted data $\boldsymbol{X}$ and the key $K \in \mathcal{K}$, in text and image data, the key rate is 
\begin{equation}
    R_K = \frac{\log(N!L!)}{N},
\end{equation}
where $N$ denotes the number of words or patches transmitted and $\boldsymbol{X} \in \mathbb{R}^{N\times L}$. In audio data, the key rate is
\begin{equation}
    R_K = \frac{\log(N!L!C!)}{NLC},
\end{equation}
where $NLC$ denotes the number of sampling points transmitted and $\boldsymbol{X} \in \mathbb{R}^{C\times N\times L}$.

\label{theorem1}
\end{theorem}

\begin{IEEEproof}
For the text and image to be transmitted, there are $N!$ possible row permutations and $L!$ possible column permutations with equal probabilities for $\boldsymbol{U}$. Therefore, the entire space for possible shared keys is $N! \cdot L!$ and it could be described by a key of length $\log_2 (N!L!)$ which is also the length of the key index in the codebook. For each patch, the average number of bits required to represent the key is equal to the total key length divided by $N$, the number of symbols to be transmitted. Similar arguments can be applied to audio data as well.
\end{IEEEproof}
We consider that the larger the key rate $R_{K}$, the more secure the encryption method is, which will be analyzed in the following section.

\begin{algorithm}[h]
	\caption{Training of the secure semantic communication framework}
	\label{algo:training}
	\KwIn{Input dataset $\mathcal{M}$, MINE network $T_{\theta}$, keyspace $\mathcal{K}$, weight factors $\alpha, \beta$,  training network $T_{Alice}, T_{Bob}, T_{Eve}$, row shuffling position $\gamma\in \{1, 2, 3\}$. }
	\KwOut{Network parameters $\theta_{A}$, $\theta_{B}$, $\theta_{E}$.}
	\While{Stop criterion is not met}
	{
		Sample data $\boldsymbol{M}$ from dataset $\mathcal{M}$.\\
		Randomly generate row shuffling key $K_{R}$ and column shuffling key $K_{C}$ from $\mathcal{K}$.\\
		\textbf{Encode with row and column shuffling:} $T_{Alice}(\boldsymbol{M},K_{R},K_{C},\gamma) \to  \boldsymbol{X}$.\\
		% \textbf{Row Shuffling:} Encryption(U) $\to$ $U_{R}$.\\
		% \textbf{Semantic Encoder:} SemEnc($U_{R}$)   $\to$ $T_{R}$
		% \textbf{Channel Encoder:} ChanEnc($T_{R}$) $\to$ $X_{R}$.\\
		% \textbf{Column shuffle:} $X_{R} \times K_{C} \to X_{K}$.\\
		Transmit $\boldsymbol{X}$ over the physical channel and receive $\boldsymbol{Y}, \boldsymbol{Z}$ respectively.\\
\liyao{Update $\theta$ by maximizing $L_{MI}^{\theta}$ of Eq.~\eqref{eqmine}.}\\
		\eIf{reveiver is Bob}
		{
			% \textbf{Colum unshuffle:} Decryption$(Y_{K},K_{C}) \to Y_{R}$.\\
			\textbf{Decode with the permutation key:} $T_{Bob}(\boldsymbol{Y},K_{R},K_{C},\gamma) \to \boldsymbol{\hat{M}}$.\\
			Compute $L_{R}(\boldsymbol{M}, \boldsymbol{\hat{M}})$.\\
			Compute $C_{S}(R_K)$ and $R_{L}$.\\	
			Update $\theta_{A}$, $\theta_{B}$ by minimizing Eq.~\eqref{finalloss}.\\
			% ChanDec($Y_{R}$) $\to$ $T_{K}$.\\
			% SemDec($T_{K}$) $\to$ $U_{K}$.\\
			% Row Decryption($U_{K}$) $\to$ $U$.\\
			% Embedding($U$) $\to$ $M'$.\\
		}{	\textbf{Decode without key:} $T_{Eve}(\boldsymbol{Z}) \to \boldsymbol{M'}$.\\
			% ChanDec($Z_{K}$) $\to$ $T_{K}$.\\
			% SemDec($T_{K}$) $\to$ $U_{K}$.\\
			Update $\theta_{E}$ by minimzing $L_{R}(\boldsymbol{M}, \boldsymbol{M'})$.\\}
	}
\end{algorithm}

\subsection{End-to-End Training}

As described in Section \ref{sec:system_model}, we represent the secure semantic communication framework in Fig.~\ref{system image} by networks of different parties, i.e., $T_{Alice}, T_{Bob}, T_{Eve}$ with parameters $ \theta_A, \theta_B, \theta_E,$ respectively. Both channel encoder and decoder are learned from end to end and are adjusted according to the data distribution to best match the channel condition. The training procedure contains two stages: the training of the legitimate channel between Alice and Bob, and the training of the wiretap channel between Alice and Eve. 
%As different data distributions may lead to inaccurate estimation of mutual information, we use the same dataset to train $T$, although this may have certain limitations in terms of generalization, it can ensure the accuracy of numerical estimation.

Letting the decoder outputs of Bob and Eve be $\boldsymbol{\hat{M}}$ and $\boldsymbol{M'}$ respectively, the message reconstruction loss function of Bob is $L_{R}(\boldsymbol{M};\boldsymbol{\hat{M}})$ which could be the cross-entropy loss in the text transmission system, or the mean square error (MSE) in the image and speech transmission system. Apart from the message reconstruction, the secrecy capacity $C_S(R_K)$ and information leakage rate $R_L$ of Eq.~\eqref{CS1} are included as losses to encourage a higher transmission rate and to prevent information leakage through the wiretap channel. Thereby, the training loss function of the legitimate channel between Alice and Bob is 
\begin{equation}\label{finalloss}
    L=L_{R}(\boldsymbol{M};\boldsymbol{\hat{M}})\ +\ \alpha R_L - \beta C_S(R_K),
\end{equation}
where $\alpha, \beta \in [0,1]$ are weight factors to balance different terms. The loss is used to optimize network parameters $\theta_A $ and $\theta_B$. \liyao{The eavesdropper Eve is assumed to be able to intercept the message being sent and train a decoder $T_{Eve}$ to reconstruct the message by minimizing the reconstruction loss
\begin{equation}
\label{eveeq}
	 L_{Eve} = L_{R}(\boldsymbol{M};\boldsymbol{M'}).
\end{equation}}
Both Bob's and Eve's communication channels are trained simultaneously in an end-to-end fashion on public datasets to establish common priors w.r.t. the message being sent. We list the training algorithm in Alg.~\ref{algo:training}. The row permutation position $\gamma \in \{1,2,3\}$ in a Transformer-based encoder-decoder due to permutation equivariance and $\gamma = 3$ in a CNN-based one. $\mathcal{M}$ is the collection of messages to be transmitted.

The network for mutual information neural estimation (MINE) is also trained end to end,
\liyao{which is used to calculate the secrecy capacity $C_S(R_K)$ and information leakage rate $R_L$ in Eq.~\eqref{CS1}.}
We minimize the loss function in Eq.~\eqref{eqmine} over $T_{\theta}$ by the update step in Line 6 of Alg.~\ref{algo:training}. The MINE network and the shuffling backbone are trained simultaneously so that the MINE network can adapt to the change in data distribution to ensure estimation accuracy.

%% file: security_analysis.tex
\section{Leakage Analysis}
\label{sec:securityanalysis}
The section inspects the security of the shuffled communication system by analyzing the privacy leakage to Eve. Hence we will analyze the channel capacity $C_{Eve}$.

\begin{figure}[ht]
    \centering
    \includegraphics[width=\columnwidth]{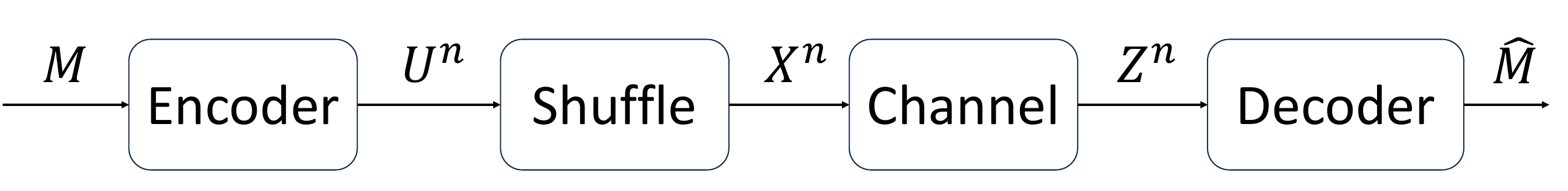}
    \caption{illustration of the communication system with shuffle operation}
    \label{simple figure}
\end{figure}

\newtheorem{myDef}{Definition}
\newtheorem{myTheo}{Theorem}

\begin{myDef}
    (Eve's Channel Capacity): The capacity of Eve's channel depicted in Fig.\ref{simple figure} is given by
    \begin{equation}
        C_{Eve} \overset{\underset{def}{}}{=} \{R_{Eve} \geqq 0: R_{Eve} \; is \; achievable \}. 
    \end{equation}
The $achievable$ meaning that $\lim\limits_{n\to\infty}P_e = 0$ where $P_e =  P(\boldsymbol{M} \neq \boldsymbol{\hat{M}} )$ is probability of error.
   
\end{myDef}
Eve's channel could be viewed as a cascade of permutation channel and the transmission channel as shown in Fig.~\ref{simple figure}, where the shuffling operation is modeled as a discrete channel. For ease of analysis, we assume the transmission channel to be a discrete memoryless channel (DMC) channel. 
The message $\boldsymbol{M} \in \mathcal{M}$ is encoded into  $\boldsymbol{U}^n \in \mathcal{U}^n$, where $n$ is the blocklength and $\boldsymbol{U}^n$ is $n$-th extension of $\boldsymbol{U}$. $\boldsymbol{X}^n \in \mathcal{X}^n$ is obtained through the permutation channel. The received message is $\boldsymbol{Z}^n \in \mathcal{Z}^n$ which the decoder decodes as $\boldsymbol{\hat{M}}$.  $\mathcal{U}$, $\mathcal{X}$, and $\mathcal{Z}$ are the collection of $U$, $X$, and $Z$, respectively. As to the convention, we define the transmission rate as $R = \frac{\log(|\mathcal{M}|)}{n}$. We have the following theorem:

\begin{theorem}
For Eve's channel, $\boldsymbol{U}^{n}-\boldsymbol{X}^{n}-\boldsymbol{Z}^{n}$ form a Markov chain, of which the information transmission rate $R_{Eve}$ is upper bounded by  
\begin{equation}
\label{rateeve}
    R_{Eve} \leq  \frac{1}{1-P_{e}} \log \left( \frac{|\mathcal{X}|(eg)^{\frac{1}{g}} }{n^{\frac{1}{g}}} \right) + o(n),
\end{equation}
where $g \in \mathbb{N_{+}}$ denotes the size of shuffle operation grain, $e$ is natural constant,  and $\frac{n}{g} \in \mathbb{N}_+$. 

\label{theorem2}
\end{theorem}

In Eq.~\eqref{rateeve}, letting $n \rightarrow \infty $ produces $R \leq 0$, and thus we have $C_{Eve} = 0$. The proof is in Supplementary Material-A.

Since the shuffle operation is completely reversible for the legitimate receiver, it guarantees that the channel capacity of the Alice-Bob communication system will not be affected by the shuffling operations. For example, Bob's channel capacity is $C_{Bob} = 1 - H(p)$ while $C_{Eve} = 0$ under the binary symmetric channel (BSC) with $p$ being the transition probability.

%% file: experiment.tex
\section{Evaluation}
\label{sec:experiment}
In this section, we evaluate the performance of our semantically secure communication system by experiments on different types of messages transmitted.

\subsection{Setup and Overhead}
\textbf{Dataset:} For text transmission, we choose the English text of the European Parliament Proceedings \cite{EuroparlData} as our dataset. Sentences of length between 4 to 30 words are selected, totaling 74,000 sentences or 1.5 million words. The dataset was split into training and testing sets with a ratio of 9:1. For image transmission, the CIFAR-10 dataset is adopted as a training dataset, which consists of 60,000 images of size $32 \times 32$ and is categorized into 10 classes. We adopt an additional dataset CIFAR-100 to verify the transferability of our system, which contains 60,000 images of size $32 \times 32$ in 100 classes. \lyr{As a representative of large image dataset, CelebA-HQ dataset \cite{celebahq} was adopted to further validate the effectiveness of the proposed method, containing 30,000 facial images of size $1024 \times 1024$, which we downsampled to $256 \times 256$ resolution due to computational resource constraints. Among these, 28,000 images were selected as the training set, while the remaining 2,000 compose the test set.} For audio data transmission, the speech dataset of Edinburgh DataShare \cite{speechdataset} is used, with \textit{clean-trainset-28psk-wav} as the training dataset and \textit{clean-testset-wav} as the test dataset. The former includes more than 10,000 \textit{.wav} files, while the latter comprises over 800 \textit{.wav} files. The sampling rate is 16kHz.

\textbf{Hyper-parameters:} The network structures and parameters of our system are listed in Table~\ref{tab:networksetting}, which is located in the Supplementary Material-B. All networks are trained with Adam optimizer and learning rate $1 \times 10^{-4}$. We assume a Gaussian channel with a Gaussian noise of variance 0.1 in the training of the encoder-decoder. The MINE network $T_{\theta}$ is implemented by MLP and is trained with Adam optimizer and learning rate $5\times 10^{-5}$. By default, the row permutation is applied to position 1 for the Transformer-based framework and position 3 for the CNN-based one. The weight factors $\alpha$ and $\beta$ are both set to 0.01.

\textbf{Baselines:} For all message types, we select three baselines, one being the conventional approach and the other being a semantic communication one. 

For text transmission, the baselines include (1) \textit{the traditional scheme} where the source coding is Huffman coding \cite{huffmancode}, and the channel coding employs Polar coding \cite{ArikanPolar} with code rate ${1 }/{2}$. According to the modulation and coding scheme (MCS) table, we select BPSK and QPSK dynamically based on different channel SNR conditions and use one-time pad as the shared key.
(2) \textit{Adversarial learning (adv)} \cite{Luo} proposed by Luo et al. is an adversarial encryption training scheme guaranteeing the accuracy of semantic communication. Since their code is not open-sourced, we re-implement their algorithm.
\liyao{(3) \textit{Learn with error (LWE) method} \cite{DeepJSCEC} proposed by Tung et al. is a training-free method based on LWE problem to guarantee the security, used for secure transmission of image data in the original paper. Since their code is not open-sourced, we re-implement their algorithm.}

For image transmission, our baselines incorporate (1) \textit{the traditional scheme} in which JPEG \cite{JPEG} is used as the source coding scheme and Polar coding with a code rate of ${1}/{2}$ is adopted as the channel coding scheme. The digital modulation scheme is QPSK and a one-time pad is used as the shared key;
(2) WITT \cite{ImageSwin20} which uses a Swin-Transformer-based network to build a semantic image transmission scheme. \liyao{(3) \textit{LWE method}.}

For the speech transmission, the baselines are (1) \textit{the traditional scheme} where an 8-bit PCM coding scheme serves as the source coding and LDPC coding \cite{LDPC1962ORIGIN} with code rate ${3}/{4}$ is employed as the channel coding scheme. The digital modulation scheme is 16-QAM and a one-time pad is used as the shared key;
(2) DeepSC-S \cite{DeepSC-S} that utilizes a CNN-based system to establish speech semantic communication, and a squeeze-and-excitation (SE) network to enhance the recovery accuracy of speech signals. \liyao{(3) \textit{LWE method}.}

% Both image and speech data security transmission systems aren't considered in the literature, so we only use the traditional scheme as the baseline and directly use the recovery accuracy of Eve to illustrate the security of our system. We assume both Bob and Eve have perfect channel state information (CSI) under fading channels.

\textbf{Metrics}: We measure the text transmission performance by the semantic error BLEU \cite{bleu} score between the original sentence of the sender and the recovered one at the receiver, and the score has been widely used in machine translation tasks. BLEU compares a pair of sentences by n-grams, i.e., a group of words, to evaluate the semantic error at varied granularities. Additionally, we adopt sentence similarity \cite{sentencebert} as another metric to evaluate the reconstruction performance. 

For image transmission, we use the Peak Signal-to-Noise Ratio (PSNR) and Multi-Scale Structural Similarity Index Measure (MS-SSIM) as the metrics to evaluate the image reconstruction performance. PSNR gauges the reconstruction quality of an image, while MS-SSIM aligns more closely with human visual perception.

The Signal-to-Distortion Ratio (SDR) is a common metric for evaluating the quality of reconstructed signals in speech transmission. Additionally, we include the Perceptual Evaluation of Speech Quality (PESQ) as an evaluation method that provides a predictive value of the subjective Mean Opinion Score (MOS) of speech quality. The PESQ value range is from -0.5 to 4.5, where a higher value indicates better reconstruction quality.

For all metrics, we expect a high value of Bob's and a low value of Eve's suggesting the semantic transmission between Alice and Bob is successful while incurring little leakage to the wiretap channel. Since the non-traditional baselines for the image and speech transmission are not designed for wiretap channels, we only evaluate the transmission performance between Alice and Bob for those methods. We assume both Bob and Eve have perfect channel state information (CSI) under fading channels.
% \begin{table}[t]
% \renewcommand\arraystretch{1.3}
% \caption{The setting of the networks.}
% \setlength{\abovecaptionskip}{1.5pt}
% \setlength{\tabcolsep}{1.3mm}
% \centering
% \label{tab:net}
% \begin{tabular}{|c|c|c|c|}
% \hline
%  & Layer Name & Units & Activation \\ \hline
% \multirow{3}{*}{Sender} & Embedding & 128 & None \\ \cline{2-4} 
%  & 4$\times$Transformer Encoder & 128 (8 heads) & Linear \\ \cline{2-4} 
%  & 2$\times$Dense & 128-16 & Relu \\ \hline
% Channel & Rayleigh/AWGN & None & None \\ \hline
% \multirow{3}{*}{\begin{tabular}[c]{@{}c@{}}Receiver\\ and \\ Eavesdropper\end{tabular}} & 2$\times$Dense & 16-128 & Relu \\ \cline{2-4} 
%  & 4$\times$Transformer Decoder & 128 (8 heads) & Linear \\ \cline{2-4} 
%  & Prediction Layer & Dictionary Size & Softmax \\ \hline
% T & 3$\times$Dense & 16-100-1 & Relu \\ \hline
% \end{tabular}
% \vspace{-0.3cm}
% \end{table}
%Part 1: 
\subsection{Overhead of Shuffling}
In our system, the additional complexity introduced by shuffling and reverse shuffling operations involves matrix multiplications. We denote the model without shuffling operation as "\emph{model-w}" and the one with shuffling operation as "\emph{model-s}". The Floating Point Operations per Second (FLOPs) of the model inference is presented in Table.~\ref{tab:complexity}. The results demonstrate that the overhead of the shuffling and reverse shuffling operations are almost negligible compared to model inference, underscoring the convenience of our method.

\begin{table}[ht!]
\renewcommand\arraystretch{1.3}
\setlength{\abovecaptionskip}{1.5pt}
\centering % 
\caption{FLOPs of different models.}
    \label{tab:complexity}
    \begin{tabular}{|l|l|l|l|}% 
        \hline
        &  model-w  &  model-s & Increased FLOPs  \\ \hline %换行
        Text  &\num{4.56e10} &\num{4.56e10} & \num{1.625e7} (0.04\%) \\\hline
        Speech & \num{ 1.997e11} & \num{2.005e11} & \num{7.16e8} (0.36\%) \\\hline
        Image  &\num{3.59e8} &\num{3.71e8} & \num{1.26e7} (3.51\%)\\ \hline
    \end{tabular}
	\vspace{-0.4cm}
\end{table}
\liyao{\subsection{Inversion Attack}
\textbf{Setting:} \lyr{We follow the setting described in Section-\ref{sec:system_model} by letting Eve run a similar reconstruction attack as the legitimate receiver (Bob), with the main difference being that Eve lacks knowledge of the data's permutation scheme employed by the transmitter.} Eve has the same training setting as Bob, which includes the training dataset, learning rate, number of training epochs, model architecture, etc., to train a decoder by Eq.~\eqref{eveeq}.}

\begin{figure}[htbp]
	\centering
	\subfloat[Text data.]{
		\includegraphics[width=0.24\textwidth]{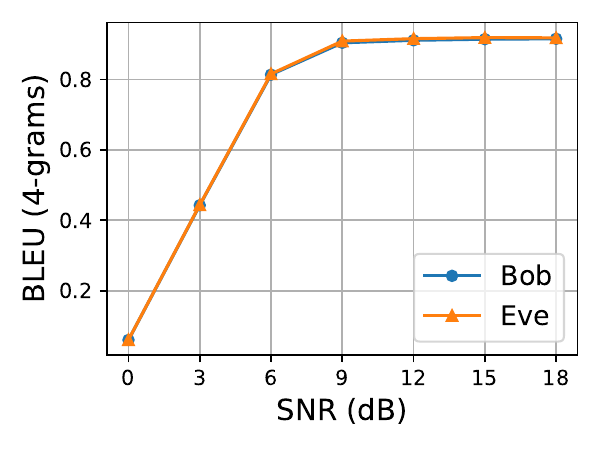}}
	\subfloat[Speech data.]{
		\includegraphics[width=0.24\textwidth]{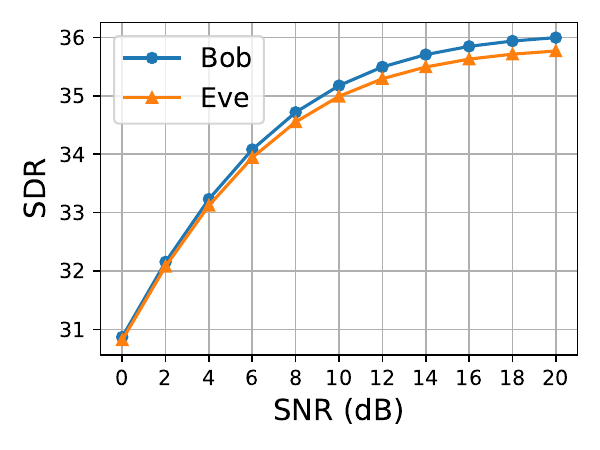}}
	\caption{\liyao{Eve's inversion attack results without protection for different types of data.}}
	\label{attack_line}
\end{figure}

\begin{figure}[h]
    \centering    \includegraphics[width=0.95\linewidth]{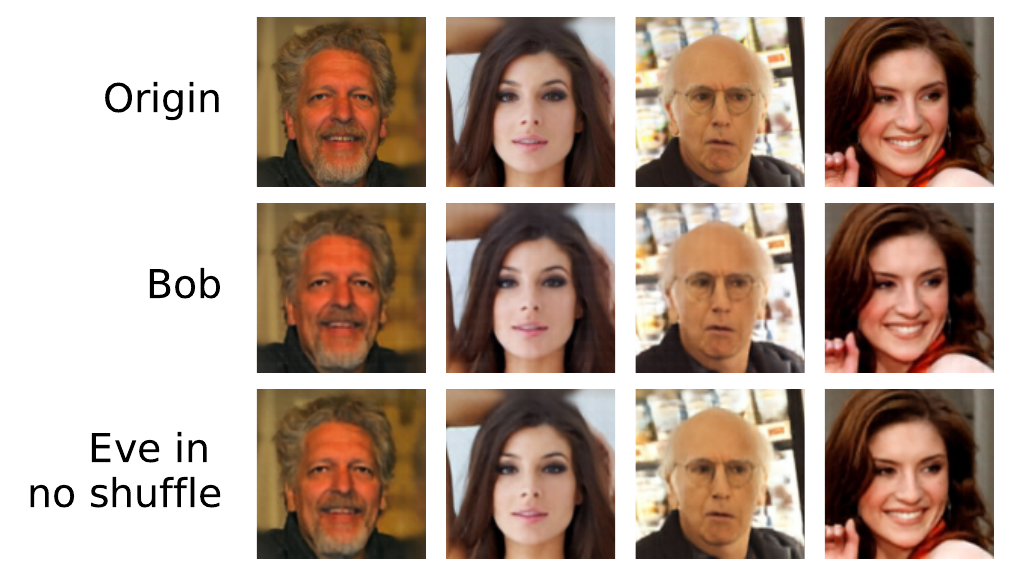}
    \caption{\lyr{CelebA-HQ Dataset: Visualization of Eve's inversion attack results without any protection. } }
    \label{attack_visual}
\end{figure}

\liyao{\textbf{Result:} Fig. \ref{attack_line} shows the results under inversion attack by Eve in the original system without applying our approach, compared with the legitimate communication of Bob's. Fig. \ref{attack_line}a is for text data and Fig. \ref{attack_line}b is for speech data. We also show the visual effects of image recovery in Fig. \ref{attack_visual}. The results reveal that Eve can achieve a similar reconstruction performance to Bob's by inversion attack, despite that Eve does not have a legitimate decoder. Hence we conclude that it is easy for an eavesdropper to acquire the sent message in the original transmission system, proving that data privacy is at risk in such a system.
}

\begin{figure*}[!htp]
	\centering
	\subfloat[Text transmission in AWGN channel.]{
		\includegraphics[width=\linewidth]{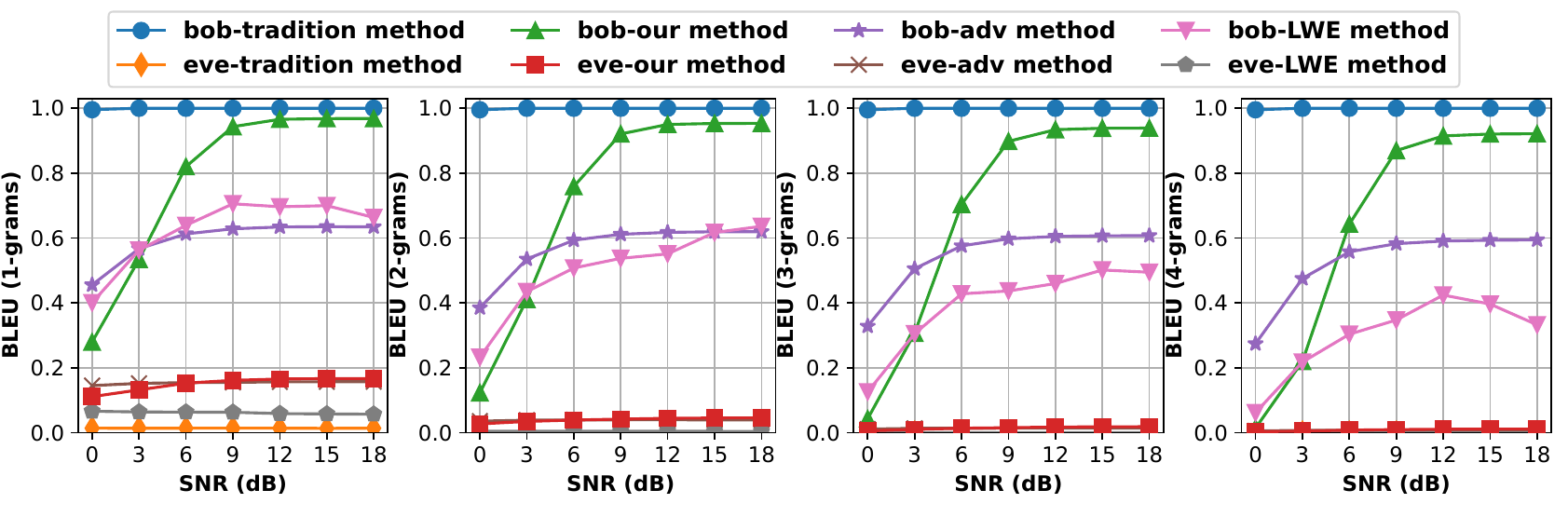}\label{awgn-bleu}}
        \hfill
	\subfloat[Text transmission in Rayleigh fading channel.]{
		\includegraphics[width=\linewidth]{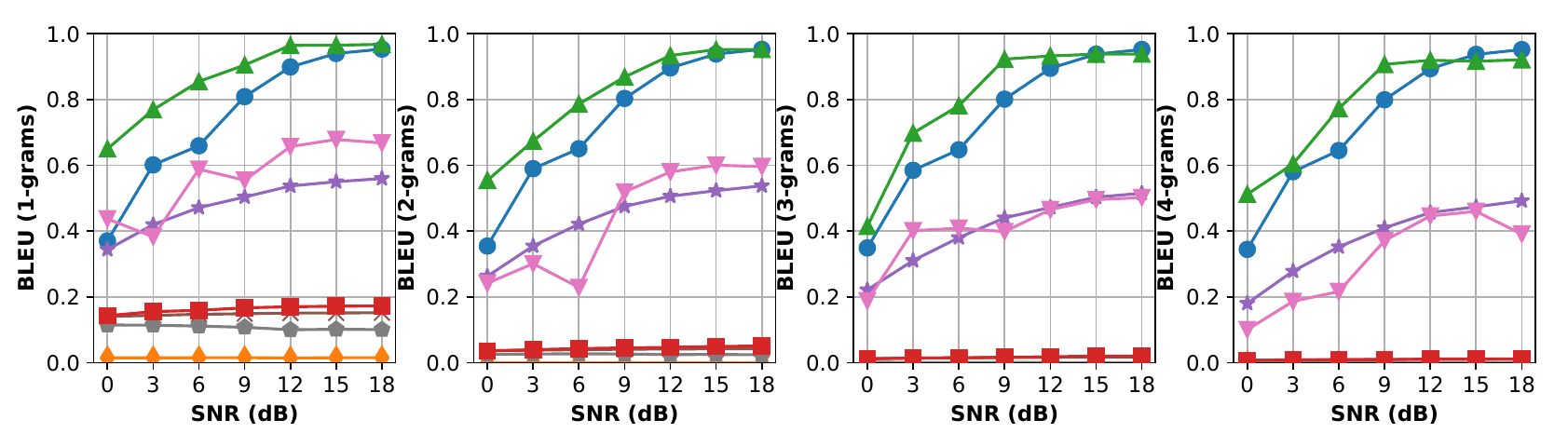}\label{ray-bleu}}
	\caption{\liyao{Bob and Eve's BLEU scores for our approach compared with the traditional method, adversarial learning-based method, and LWE method in (a) AWGN channel, and (b) Rayleigh fading channel.}}
	\label{BLEU}
    % \vspace{-0.4cm}
\end{figure*}

\liyao{In the following sections, we show Eve's reconstruction results by data types under our approach and other baselines. }

\subsection{Text Reconstruction Results}

Fig.~\ref{awgn-bleu} and \ref{ray-bleu} illustrate the BLEU score of Bob and Eve under different signal-to-noise ratios (SNRs) in both AWGN and Rayleigh channels.
Note that the traditional method uses BPSK as the digital modulation scheme when the channel SNR is below 6dB according to the MCS table specifications, which is the most robust digital modulation scheme. The traditional method performs best in AWGN but is poorer than our method in Rayleigh channels. This is because the traditional method aims to recover every bit, posing strict requirements on channel conditions. In contrast, our method achieves relatively high BLEU in both AWGN and Rayleigh channels for Bob, demonstrating robustness against the fading channel. Adv method has degraded performance under two channel circumstances, mostly due to the difficulty in learning encryption with neural networks. \liyao{LWE method achieves poor performance, especially in 4 grams, which may be because the method introduces error terms in encryption, resulting in the largest decoding error. Since the greedy decoding algorithm is used for text data, an error in one word would cause the entire sentence to fail to decode, thus leading to the worst performance. However, LWE is more robust than Adv in the Rayleigh channel, as the training-free approach does not rely on the specific channel state and thus it could maintain robustness even if the state drastically changes from that of training, particularly with fading channels.}

% \begin{figure}[ht!]
% 	\centering
% 	\subfloat[Text transmission in AWGN channel.]{
% 		\includegraphics[width=0.47\columnwidth]{Figure/AWGN_Sentence_Similarity.pdf}\label{awgn-simili}}
%         \hfill
% 	\subfloat[Text transmission in Rayleigh fading channel.]{
% 		\includegraphics[width=0.47\columnwidth]{Figure/Rayleigh_Sentence_Similarity.pdf}\label{ray-simili}}
% 	\caption{Bob and Eve’s sentence similarity for the three methods in (a) AWGN channel, and (b) Rayleigh fading channel.}
% 	\label{sentence-similarity}
% 	\vspace{-0.3cm}
% \end{figure}

Fig.~\ref{BLEU} also presents Eve's BLEU score under different SNRs. While the traditional method employs a one-time pad which affords perfect secrecy, the score of Eve is the lowest among all. Our method and adv method share similar close-to-zero BLEU except for the 1-grams, demonstrating a strong capability of hiding semantic information from Eve. \liyao{LWE method also demonstrates similar results with all BLEUs close to zero.} As we analyze, Eve's BLEU 1-grams are not close to zero mostly because special tokens are more likely to be recovered by random reconstruction. The evidence that the recovered results of Eve do not improve with the increase of SNR further confirms that Eve's reconstruction is subject to random guesses. The results under the sentence similarity metrics are presented in Supplementary Material-D.
% \ref{sec:sentence_simi}
% \begin{figure}[htbp]
% 	\centering
% 		\includegraphics[width=0.99\columnwidth]{Figure/Sentence_Similarity.pdf}
% 	\caption{\liyao{Bob and Eve’s sentence similarity for different methods in (a) AWGN channel, and (b) Rayleigh fading channel.}}
% 	\label{sentence-similarity}
% 	\vspace{-0.3cm}
% \end{figure}
% Fig.~\ref{sentence-similarity} displays the sentence similarity scores of four methods. The results mostly agree with that under BLEU. In particular, our method achieves a sentence similarity score over 0.8 at SNR greater than 6 dB in the Rayleigh channel, indicating a successful recovery of semantic information at the sentence level. 
For a more intuitive understanding of our results, we sample some recovered sentences in Table~\ref{tab:sentence}, which shows that Bob can accurately restore the original sentence, while Eve's recovery results share no common semantic information with the original sentence. The recovered sentences appear to make sense mostly due to the use of greedy decoding which ensures that the context conforms to some specifications.

\begin{table}[htb]
    \renewcommand\arraystretch{1.2}
        \centering
        \caption{The sentences decoded by Bob and Eve in AWGN channel, SNR$=9$dB.}
        \begin{tabular}{|c|p{6cm}|}
        \hline
           Alice's message  &   thank you very much for your debate  i really believe we will come back for the debate and agree on the content of a european energy policy\\ \hline
            Bob's decoding &  thank you very much for your debate  i really believe we will come back for the debate and agree on the content of a european energy policy \\ \hline
            Eve's decoding & i am delighted that the european parliament has decided to vote on the report and i hope that the commission will be able to accept the following amendments  \\ \hline
        \end{tabular}
        \label{tab:sentence}
        % \vspace{-0.2cm}
    \end{table}

% \begin{table*}[t!]
%     \renewcommand\arraystretch{1.2}
%         \centering
%         \caption{The sentences decoded by Bob and Eve in AWGN channel, SNR$=15$dB.}
%         \begin{tabular}{|c|p{11cm}|}
%         \hline
%            Alice's message  &   i hope that the report will have a good result and will help everyone without exceptions\\ \hline
%             Bob's decoding &  i hope that the report will have a good result and will help everyone without exceptions \\ \hline
%             Eve's decoding & the next item is the commission statement on the situation in the middle east  \\ \hline
%         \end{tabular}
%         \label{tab:sentence}
%         \vspace{-0.4cm}
%     \end{table*}

\begin{figure}[htbp]
    \centering
    \includegraphics[width=0.9\linewidth]{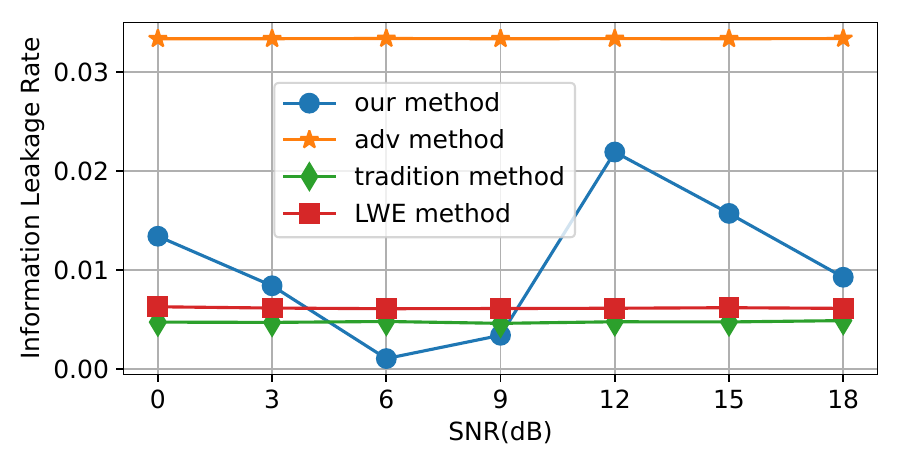}
    \caption{\liyao{Information Leakage Rates $R_L$ of different methods under AWGN channel.}}
    \label{RL}
\end{figure}

For a deeper analysis, we depict the magnitude of information leakage rate by mutual information $R_L$ in Fig.~\ref{RL}, which is calculated by the pre-trained MINE network. It should be noted that the maximal mutual information is 1 bit/channel use. The results are consistent with those under BLEU and sentence similarity that, the traditional method achieves the lowest $R_L$ due to the use of one-time pad and our method enjoys a sufficiently low leakage rate. 

% \begin{figure}[htbp]
%     \begin{minipage}{0.48\linewidth}
%     \centering
%     \includegraphics[width = 0.99\linewidth]{Figure/Rateleakage.pdf}
%     \caption{Information Leakage Rates of three methods (AWGN). }
%     % \strut
%     \label{RL}
%     \end{minipage}
%     \hfill
%     \begin{minipage}{0.48\linewidth}
%     \centering
%     \includegraphics[width = 0.99\linewidth]{Figure/image/AWGN_PSNR.pdf}
%     \caption{PSNR scores of Bob and Eve in different shuffling modes.}
%     \label{keyspace}
%     \end{minipage}
% \end{figure}

% \begin{figure}[htbp]
% 	\centering
% 	\subfloat[AWGN channel.]{
% 		\includegraphics[width=0.24\textwidth]{Figure/image/PSNR_AWGN.pdf}}
% 	\subfloat[Rayleigh fading channel.]{
% 		\includegraphics[width=0.24\textwidth]{Figure/image/PSNR_Rayleigh.pdf}}
% 	\caption{Image transmission \lyr{on CIFAR-100}: Bob and Eve’s PSNR scores for different methods in (a) AWGN channel and (b) Rayleigh channel.}
%     \label{fig:psnr}
% \end{figure}

\subsection{Image Reconstruction Results}

\begin{figure}[htbp]
	\centering
	\subfloat[AWGN channel.]{
		\includegraphics[width=0.24\textwidth]{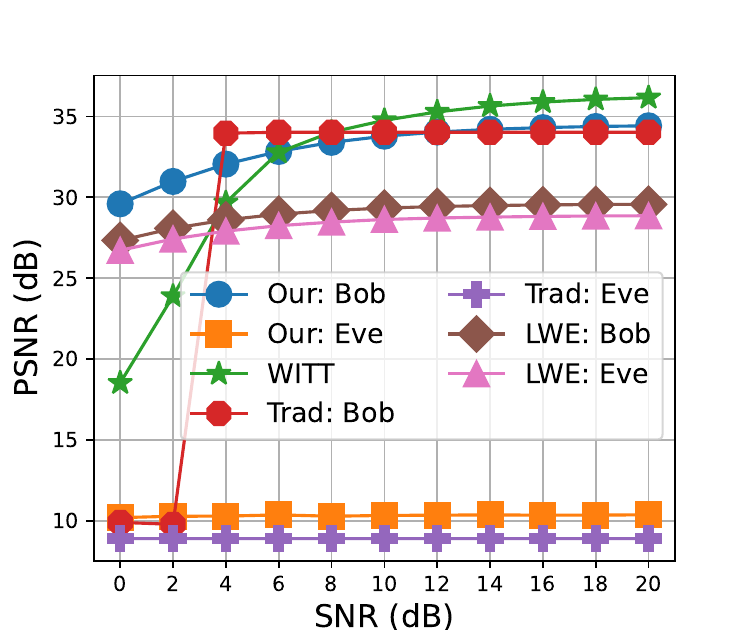}}
	\subfloat[Rayleigh fading channel.]{
		\includegraphics[width=0.24\textwidth]{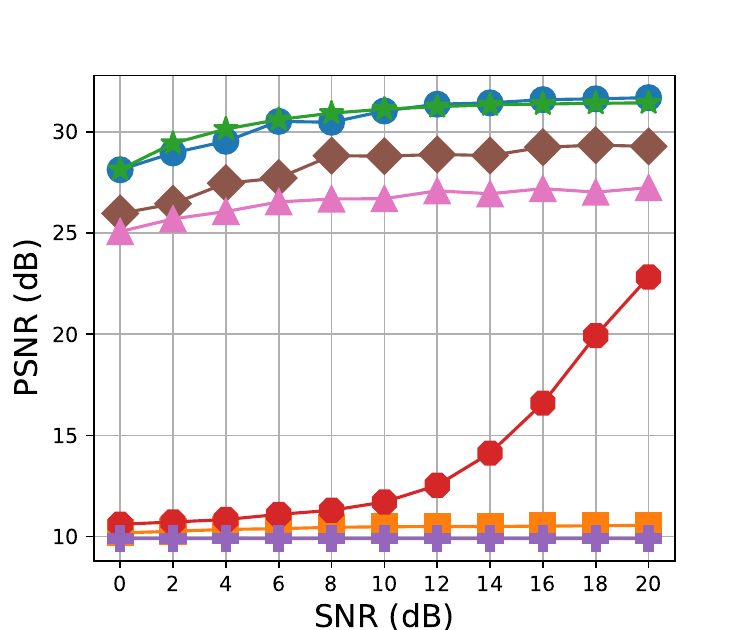}}
	\caption{\lyr{Image transmission on CelebA-HQ: Bob and Eve’s PSNR scores for different methods in (a) AWGN channel and (b) Rayleigh channel.}}
    \label{fig:psnr}
\end{figure}

\lyr{Fig.~\ref{fig:psnr} reports the PSNR scores on CelebA-HQ under AWGN and Rayleigh channels. Our approach demonstrates a superior performance to the baseline across both channels, particularly in Rayleigh channels, showing the capability of our system to adjust the input distribution according to the channel conditions. The traditional method (Trad) exhibits poor results in the Rayleigh channels, primarily because JPEG is a lossy compression method and struggles to yield adequate results even in the optimal channel condition. WITT is a competitive method compared to ours, but still has a suboptimal performance in AWGN channels under low SNRs. The performance of LWE is consistent with the results in the original paper but is still inferior to ours due to the error introduced.}

Note that DNN-based methods are in general more robust against noise than traditional methods, especially in low SNRs, and we consider this due to the end-to-end training adjusting the encoder and decoder according to the channel conditions and data distributions. In contrast, since the encoder and decoder in the traditional method are hand-crafted, it is more vulnerable to the impact of channel noise and fading. \lyr{Meanwhile, Eve's PSNR score is below 11dB under all SNRs by our method,} close to the results of the traditional method, indicating poor reconstruction image quality of the eavesdropper. LWE method is not as secure as stated in its original paper, which may be because Eve's capabilities and training resources in our setup exceed those in \cite{DeepJSCEC}, allowing Eve to eventually break the defense. The results under the MS-SSIM metrics are presented in Supplementary Material-E.
% Appendix \ref{sec:msssim}.

\subsection{Speech Transmission Results}
\begin{figure}[htbp]
	% \vspace{-0.6cm}
	\centering
	\subfloat[AWGN channel.]{
		\includegraphics[width=0.24\textwidth]{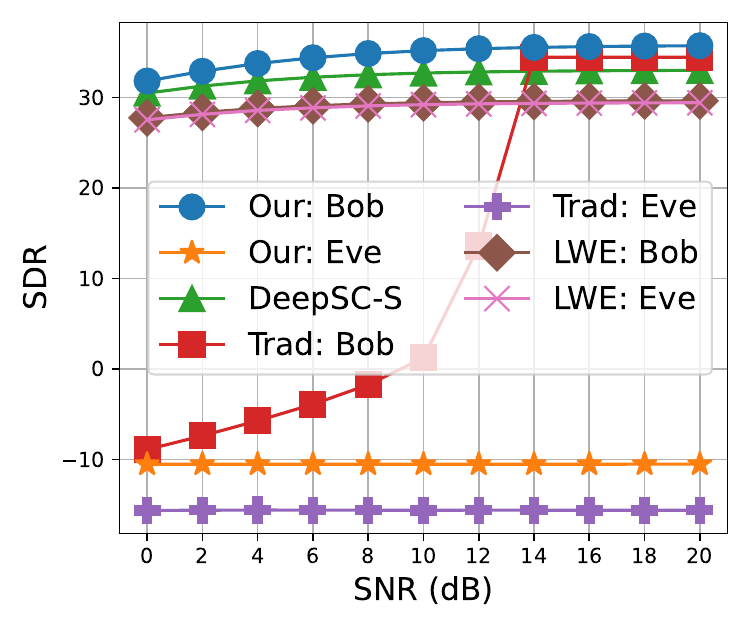}\label{SDR_AWGN}}
        \subfloat[Rayleigh fading channel.]{
        \includegraphics[width=0.24\textwidth]{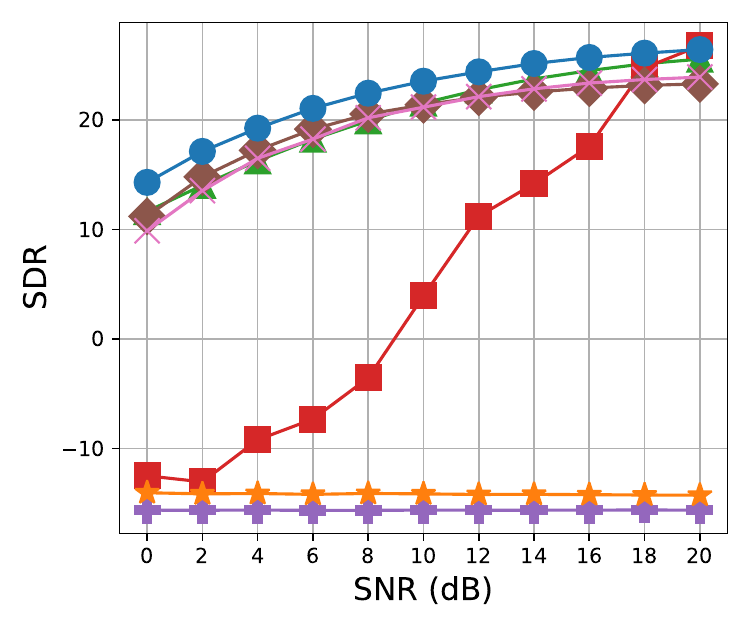}\label{SDR_Rayleigh}}
	% \vspace{-0.3cm}
	\caption{\liyao{Speech transmission: Bob and Eve’s SDR score for different methods in (a) AWGN channel and (b) Rayleigh channel.}}
	\label{SDRscore}
\end{figure}

Fig.~\ref{SDRscore} illustrates the SDR performance versus SNRs in AWGN and Rayleigh channels. Our method demonstrates the optimal performance and DeepSC-S has a close performance to ours, affirming that the shuffling operations do not compromise the reconstruction accuracy of speech data. Compared to the semantic approach, the rise in SDR with the increase of SNR is steeper for the traditional method, since the digital modulation scheme 16QAM is error-prone at low SNRs. Although the reconstruction accuracy can be improved by using simpler modulation methods than 16QAM, the bandwidth efficiency would be reduced as a sacrifice. The sharp rise may also be attributed to the use of 8-bit PCM in source encoding and the silent part of the speech signal is also susceptible to noise interference, both of which reduce the reconstruction accuracy. \liyao{LWE method is inferior to other methods under the AWGN channel, but it shows superior performance under the Rayleigh channel, which again confirms that the train-free method is more robust in the fading channel.}

%origin-version
% Fig.~\ref{SDRscore} illustrates the SDR performance versus SNRs in AWGN, Rayleigh, and Rician channels. Our method demonstrates the optimal performance and DeepSC-S has a close performance to ours, affirming that the shuffling operations do not compromise the reconstruction accuracy of speech data. Compared to the semantic approach, the rise in SDR with the increase of SNR is steeper for the traditional method, since the digital modulation scheme 16QAM is error-prone at low SNRs. Although the reconstruction accuracy can be improved by using simpler modulation methods than 16QAM, the bandwidth efficiency would be reduced as a sacrifice. The sharp rise may also be attributed to the use of 8-bit PCM in source encoding and the silent part of the speech signal is also susceptible to noise interference, both of which reduce the reconstruction accuracy. 

\liyao{Note that semantic-based approaches, i.e., our method, DeepSC-S, and LWE, are quite robust in the low SNR region}, especially in the AWGN channel, which may be due to the redundancy in the speech signal and thus the noise has less impact on the informative part. Fig.~\ref{SDRscore} also presents Eve's SDR results of our method and traditional method, which are consistently below 0 dB, suggesting that Eve cannot recover any useful information. \liyao{In contrast, in LWE, Eve has a performance almost identical to Bob's, indicating failure to defeat the eavesdropping attack. We analyze that it may be because the error spread is limited in CNN networks compared to Transformers, leading to a more successful attack by Eve. } The results under the PESQ metrics are presented in Supplementary Material-F.

In conclusion, the experiments on three types of data transmission show the superiority of our shuffling-based method in recovering the message received by the legal receiver while preventing leakage to the eavesdropper.

\subsection{Ablation Study}

\begin{table}[htbp]
    \renewcommand\arraystretch{1.3}
    \caption{The $R_{K}$ in different shuffling modes.}
    \setlength{\abovecaptionskip}{1.5pt}
    \setlength{\tabcolsep}{1.3mm}
    \centering
        \label{tab:RK}
        \begin{tabular}{|l|l|l|l|l|}
        \hline
         & Row & Column & Row and Column & 3-order \\ \hline
        Text & 3.63 & 1.42 & 5.05 & / \\ \hline
        Image & 4.62 & 0.69 & 5.31 & / \\ \hline
        Speech & \num{9e-4} & \num{9e-4} & \num{1.8e-3} & \num{7e-3} \\ \hline
        \end{tabular}
\end{table}

\subsubsection{Different key rates} We vary the key rates by using different permutation modes. Table~\ref{tab:RK} shows $R_{K}$s under different shuffling modes --- row shuffling, column shuffling, and 3-order shuffling where row, column, and channel shuffling are performed together. The key rate is higher when both row and column shuffling are performed together than when a single type of shuffling is performed. 

%删改1-图片
% \begin{figure}[htbp]
%     \centering
%     \includegraphics[width = 0.8\linewidth]{Figure/keyspace.pdf}
%     \caption{Visualization of Eve's received images. The first row is the origin image, the second row is Bob's recovery results, and the third to sixth are Eve's recovery results under different shuffling modes.}
%     \label{visual}
% \end{figure}

%删改1：删掉图片数据下，eve-bob的可视化结果。
%(origin) To see how secure it is by different key rates, we show Eve's PSNR and the visualization effect of its reconstructed images in both Fig.~\ref{keyspace} and Fig.~\ref{visual}. Fig.~\ref{keyspace} reveals that Eve can achieve PSNR scores close to Bob's in the non-shuffling case, emphasizing how easy it is for an eavesdropper to acquire the message sent in the non-protected scene. The PSNR is lowest when both row and column shuffling are performed, i.e., at the highest key rate. The results are also consistent with the reconstruction visualization effect in Fig.~\ref{visual} where silhouette vaguely appears in column shuffling accompanied by noise while pure noise is recovered from the row-column shuffling. The results support that shuffling indeed serves as an effective key in protecting the semantic secrecy of the transmitted messages, and the higher the key rate, the better the protection.

\begin{figure}[htbp]
    \centering
    \includegraphics[width = 0.85\linewidth]{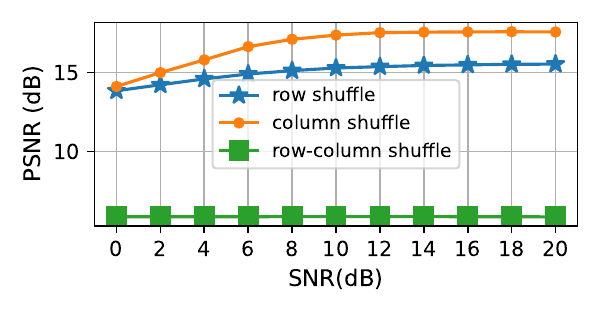}
    \caption{\liyao{Image transmission on CIFAR-100: PSNR scores of Eve in different shuffling modes.}}
    \label{keyspace_line}
\end{figure}

\begin{figure}[htbp]
    \centering
    \includegraphics[width = 0.99\linewidth]{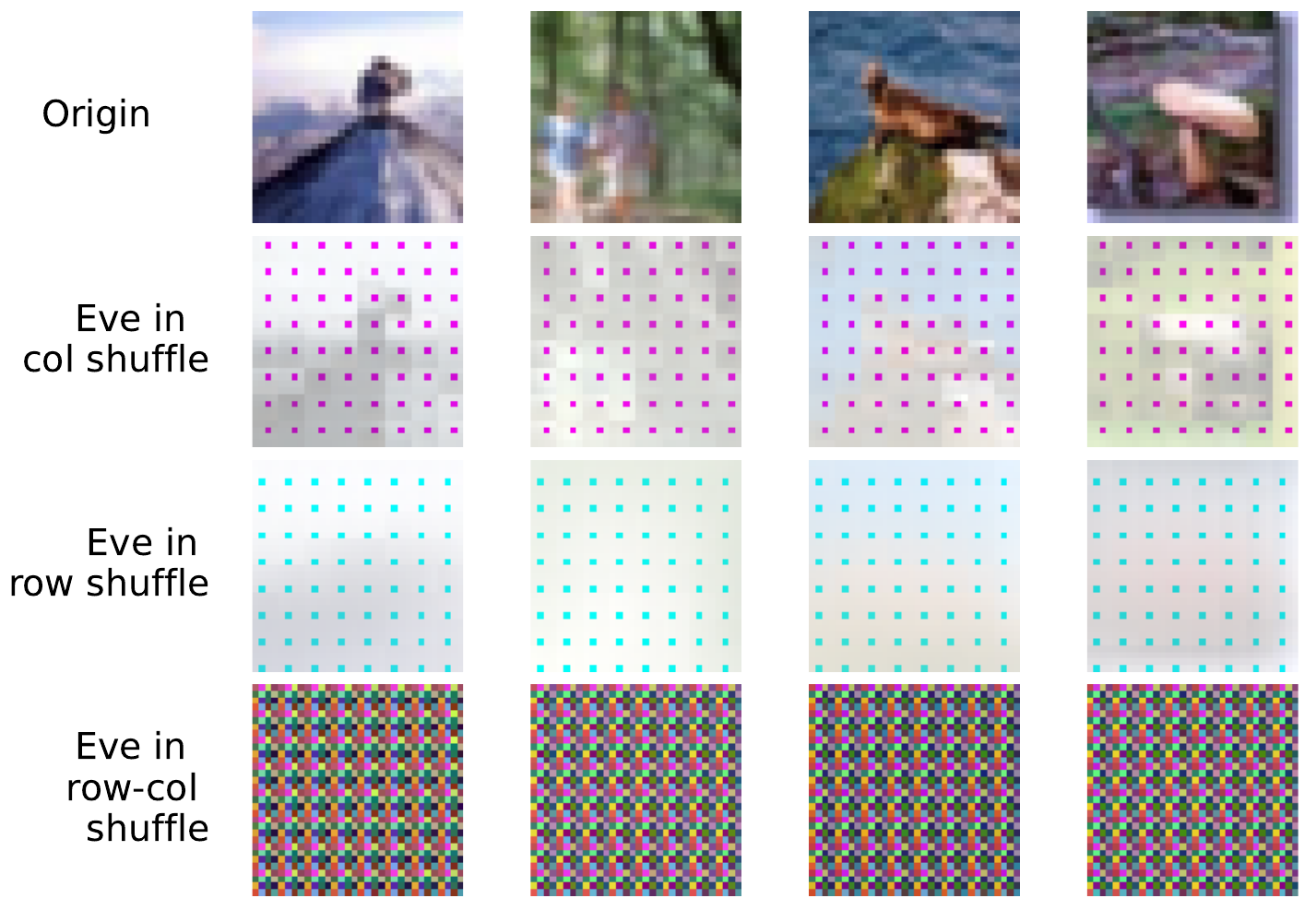}
    \caption{\liyao{Image transmission on CIFAR-100: visualization of Eve's reconstruction in different shuffling modes.}}
    \label{keyspace}
\end{figure}

\textbf{Image Data Result:} We show Eve's PSNR under different SNRs in Fig.~\ref{keyspace_line} and visualize the results in Fig.~\ref{keyspace} to see how secure it is by different key rates. The PSNR is lowest when both row and column shuffling are performed, i.e., at the highest key rate. The results support that shuffling indeed serves as an effective key in protecting the semantic secrecy of the transmitted messages, and the higher the key rate, the better the protection.

% \begin{figure}[htbp]
%     \centering
%     \includegraphics[width = 0.85\linewidth]{Figure/image/celeba/keyspace_line_celeba.pdf}
%     \caption{\lyr{Image transmission on CelebA-HQ: PSNR scores of Eve in different shuffling modes.}}
%     \label{keyspace_line}
% \end{figure}
% \begin{figure}[htbp]
%     \centering
%     \includegraphics[width = 0.99\linewidth]{Figure/image/celeba/keyspace_celebahq.pdf}
%     \caption{\lyr{Image transmission on CelebA-HQ: visualization of Eve's reconstruction in different shuffling modes.}}
%     \label{keyspace}
% \end{figure}


\liyao{Fig. ~\ref{keyspace} demonstrates that, under pure column shuffling, Eve can approximately reconstruct the color distribution of the image. This is because column shuffling destroys the spatial relationship within each ViT patch but retains the relationship between patches. With row shuffling, this relationship between patches is broken. When both row and column shuffling are applied,  Eve fails to recover any meaningful image information, as the spatial relationship within an image is destroyed. We consider the regular patterns exhibited in row-column shuffling because the ViT model operates on a basic unit of 4$\times$4 patch.}
%%%%%%%%
% When only row shuffling is applied, Eve can roughly recover the general outline of the original information with relatively smooth results, though some details are missing. For instance, in the third column results, while the original face belongs to an elderly person, the reconstructed output appears as middle-aged. This occurs because row shuffling operates between different patches, allowing Eve only to assemble them into a seemingly plausible facial image. However, due to the lack of relative positional information between patches, only the approximate contour can be recovered.

% When only column shuffling is applied, Eve's reconstruction results more closely resemble the original image, though each recovered patch appears coarser. This is because column shuffling operates within individual patches, preventing Eve from accurately restoring the texture information of each image block.

% When both row and column shuffling are simultaneously employed, Eve can only recover a face-like image. This limited success stems solely from Eve's complete access to the training dataset (knowing that faces are being transmitted), yet still fails to reveal any meaningful information about the original image. These results conclusively demonstrate the security effectiveness of the combined row-column shuffling operation.
%%%%%%%%

\begin{figure}[htb]
	% \vspace{-0.4cm}
	\centering
	\subfloat[Non-shuffling]{
		\includegraphics[width=0.49\linewidth]{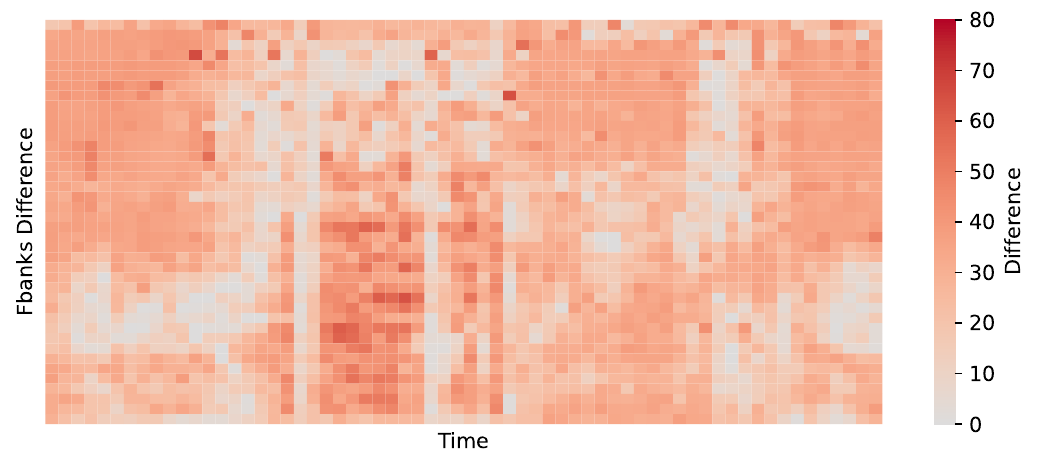}\label{noshuffle}}
    \subfloat[Row shuffling]{
        \includegraphics[width=0.49\linewidth]{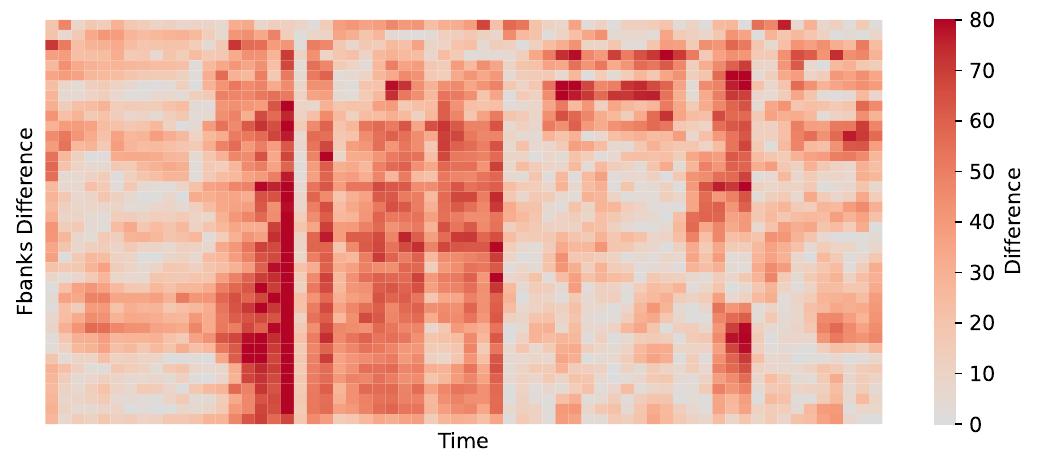}\label{rowshuffle}}\\
        % \qquad
	%让图片换行
    \subfloat[Channel shuffling]{
        \includegraphics[width=0.49\linewidth]{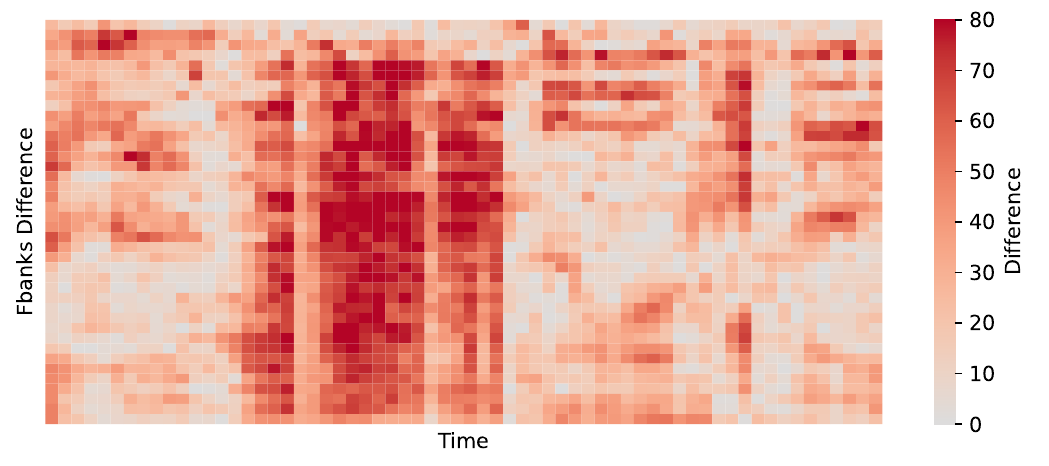}\label{channelshuffle}}
    \subfloat[3D shuffling]{
        \includegraphics[width=0.49\linewidth]{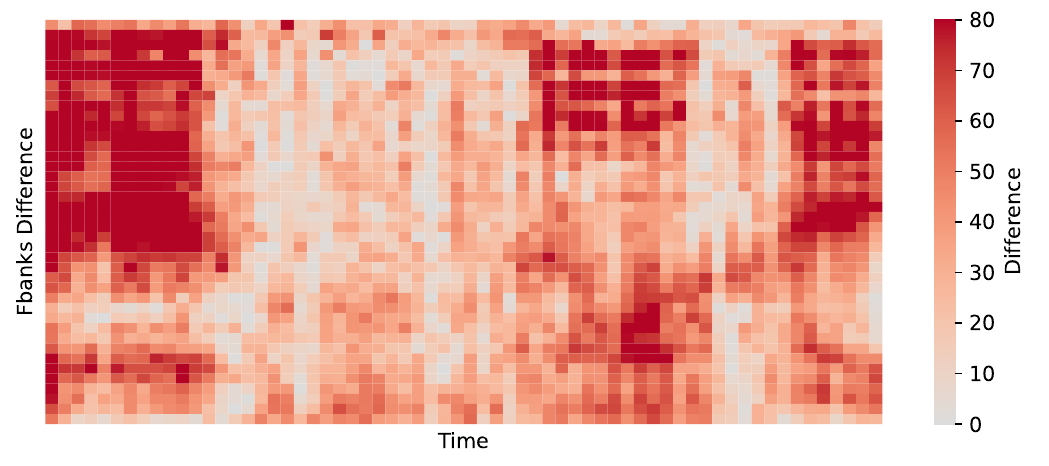}\label{3dshuffle}}
	% \vspace{-0.3cm}
	\caption{Speech transmission: Eve's spectrogram differential heat map in different shuffling modes.}
	\label{speechshuffle}
\end{figure}

\textbf{Speech Data Result:} Fig.~\ref{speechshuffle} depicts the spectrogram differential heat map of the speech signal in different shuffling modes. The spectrogram differential heat map is calculated by the difference between the spectrogram of the original speech signal and the spectrogram of the recovered speech signal, the darker the color, the larger the difference. The spectrogram differential heat map of the speech signal in the absence of shuffling is almost light everywhere, indicating that Eve can accurately recover the original speech signal. With an increasing key rate, the heat map gradually darkens, suggesting harder reconstruction by Eve. This observation demonstrates that shuffling can effectively enhance the secrecy of the CNN-based semantic communication system.

%text result part
\begin{figure}
    \centering
    \includegraphics[width=0.99\linewidth]{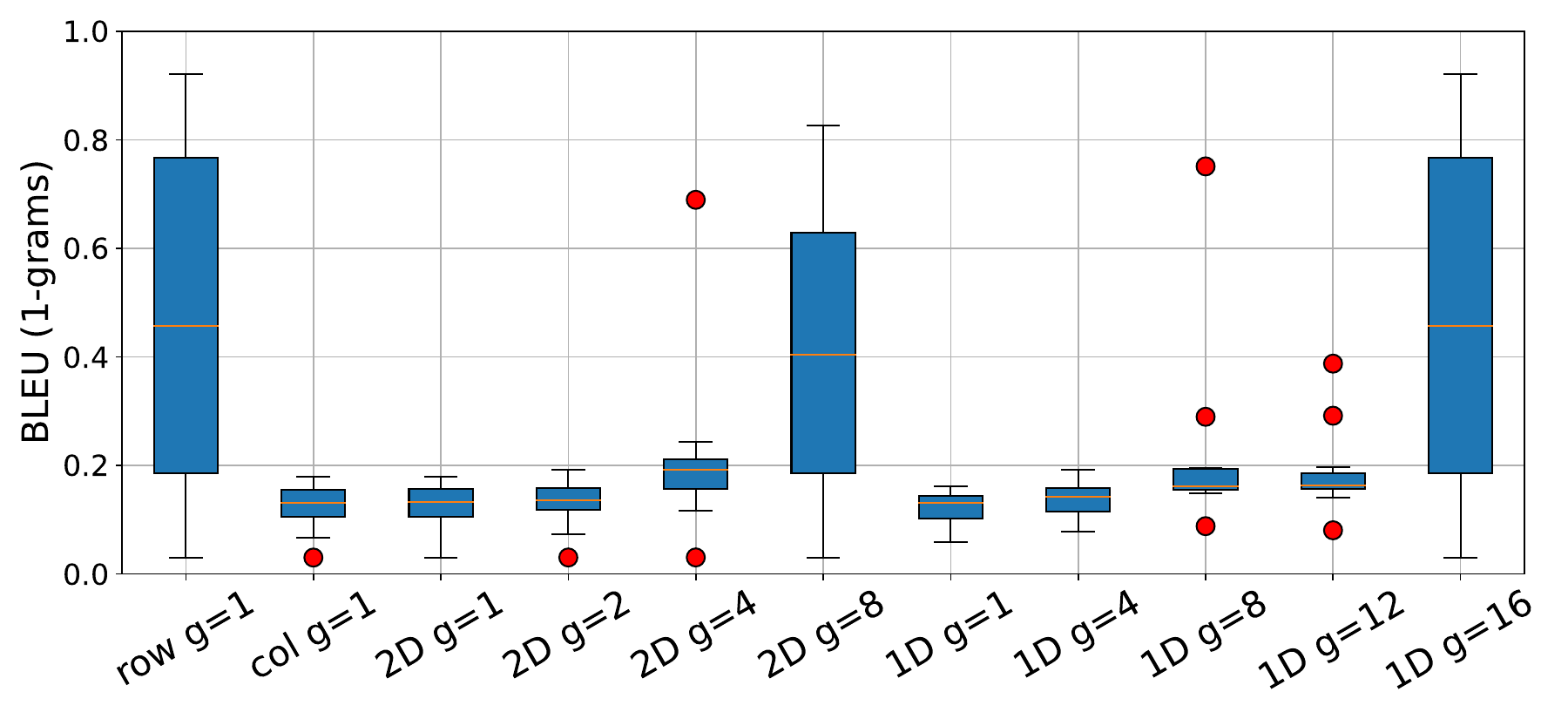}
    \caption{Text transmission: Eve's BLEU score (1-grams) with different key rates. The `g' represents `grain' which is the shuffling granularity. The smaller the granularity, the higher the key rate.}
    \label{textboxfigure}
\end{figure}
\textbf{Text Data Result:} We also study how key rates affect the secrecy in text transmission. Fig.~\ref{textboxfigure} illustrates Eve's BLEU score (1-grams) under different key rates realized by different shuffling grains. For example, \emph{"g=2"} means that every two elements are bound as one basic shuffling unit. \emph{2D} denotes the simultaneous row and column shuffling on the matrix-valued feature. \emph{1D} means applying shuffling on a one-dimensional sequence flattened from the original feature. Notably, the key rate of 1D is larger than that of 2D under the same grain. Eve trains a decoder model for each shuffling mode respectively and tries all decoder models to decode the correspondingly shuffled message. 

The recovery results are presented in boxplots. The secrecy under row shuffling is better protected than that by column shuffling as the key rate of the former is larger.

We have a theoretical direction that the larger the keyspace, the better the security performance of the system, Fig.~\ref{textboxfigure} follows this direction, but the following exceptions occur: Firstly, the result shows that the security performance of \emph{"row"} is better than \emph{"col"} whereas the keyspace size of \emph{"row"} is smaller. 

We think this is because the global attention mechanism used is the Transformer which might guarantee that Eve can use the information between each token to get better recovery results without knowing the location information. What's more, each token in the text transmission is a complete word, and Fig.~\ref{BLEU} shows that there are many single-character words in the dataset, which may also contribute to the above results. An exception occurs where the reconstruction of `1D g=12' of a smaller key rate is slightly worse than `1D g=8.' It may be because shuffling by `g=8' may lead to a more straightforward reconstruction as the dimension of each token from the channel encoder is 16. In summary, the shuffling operation achieves secrecy by altering the original feature distribution, and the greater the distributional change leads to better secrecy.

\subsubsection{Shuffling Position Options}
\begin{figure}[htbp]
    % \vspace{-0.4cm}
    \centering
    \includegraphics[width=0.99\linewidth]{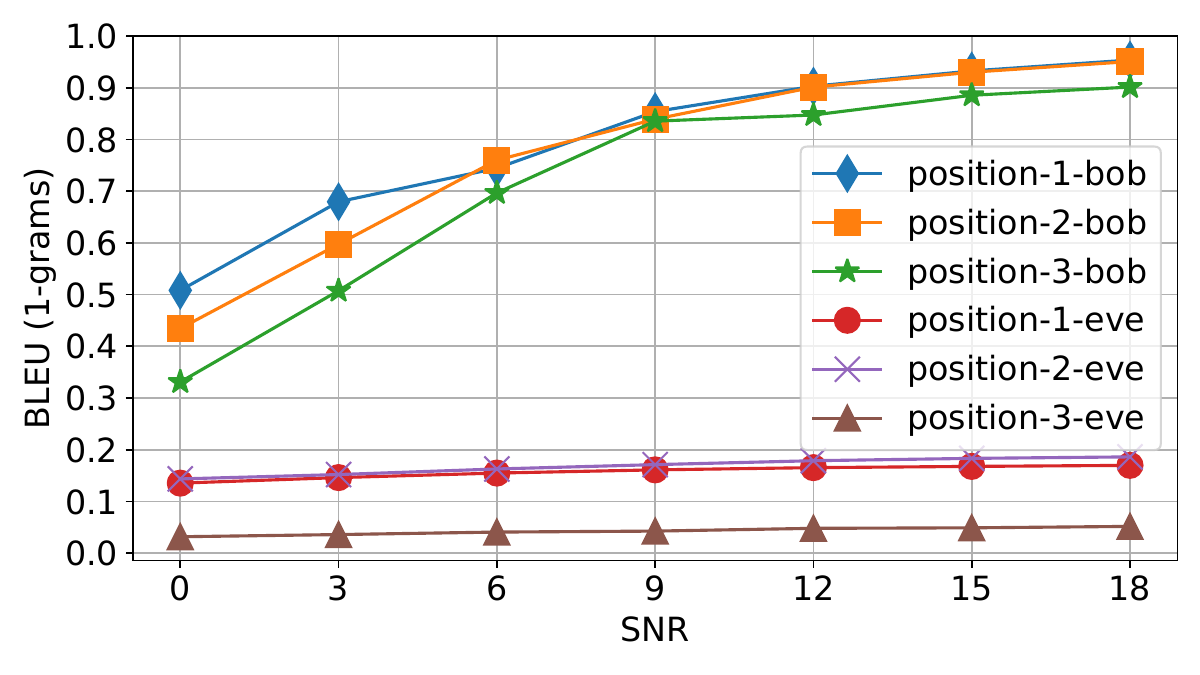}
    \caption{How different row shuffling positions affect Bob's and Eve's reconstruction performance in text transmission. }
    \label{position}
\end{figure}

In the Transformer-based semantic communication system, the row shuffling operation could be placed in three positions due to the permutation equivariance of the model. We investigate the impact of different shuffling options on the system. Fig. \ref{position} shows Bob's and Eve's BLEU (1-grams) scores for the three shuffling positions. Bob achieves over $0.9$ BLEU score in all three at the high SNR regime, verifying the three shuffling positions are almost identical to the legitimate receiver. In the low SNR regime, the performance from best to worst are positions 1, 2, and 3. Eve obtained low BLEU scores across all SNRs in all three positions, with the lowest at position 3, indicating negligible leakage to the eavesdropper.

This may be because, at position 3, row shuffling is performed to the output of the sender's joint source-channel coding, so that only the lower part of the network (from position 3 to final loss) tries to adapt to the shuffling, leaving the upper part not aware of shuffling. In contrast, if the shuffling takes place at position 1, the entire network would adapt to shuffling by adjusting the data distribution, and with more parameters, the decoding accuracy is higher. The point is further confirmed by the comparison between positions 1 and 2. Eve, without the shared key, needs to learn reversing shuffling in decoding, which is a harder task when more parameters are involved in the shuffling adaptation. Nevertheless, all three shuffling positions are favorable depending on the specific accuracy-leakage tradeoff required.

\liyao{
\subsubsection{Choices of $\alpha$ and $\beta$}
\label{sec:ablation}
In this section, we examine the impact of different values of $\alpha$ and $\beta$ on the system performance, which are weight factors for the information leakage rate $R_L$ and the secrecy capacity $C_S$, respectively, in Eq.~\eqref{finalloss}. Performance is evaluated in text data using the BLEU (2-grams) metric.}
\begin{figure}[htb]
	\centering
	\subfloat[Bob.]{
		\includegraphics[width=0.24\textwidth]{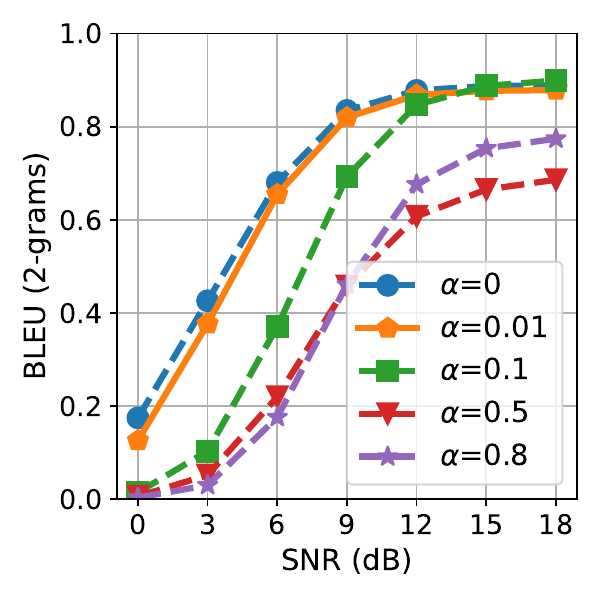}}
        \subfloat[Eve.]{
        \includegraphics[width=0.24\textwidth]{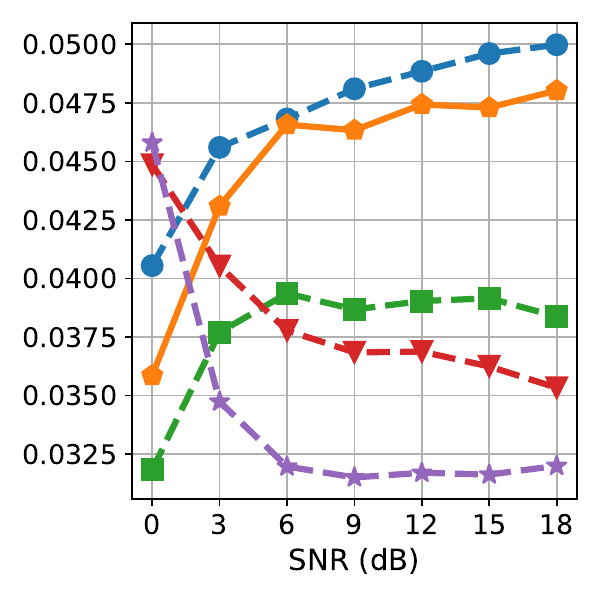}}
	\caption{\liyao{Bob's and Eve's BLEUs at $\beta$=0.01 and varying $\alpha$s. }}
	\label{fig:alpha}
\end{figure}

\liyao{
The results of fixing $\beta=0.01$ and varying the value of $\alpha$ are shown in Fig. ~\ref{fig:alpha}, where the solid yellow line denotes the results with default $\alpha = 0.01$. It can be observed that since $\alpha$ controls the weight of $R_L$, Eve's performance is heavily impacted. An increase in $\alpha$ leads to a lower BLEU of Eve, indicating less leakage. As a side effect, Bob's BLEU also declines but remains orders of magnitude higher than Eve's.}

\begin{figure}[htb]
	\centering
	\subfloat[Bob.]{
		\includegraphics[width=0.24\textwidth]{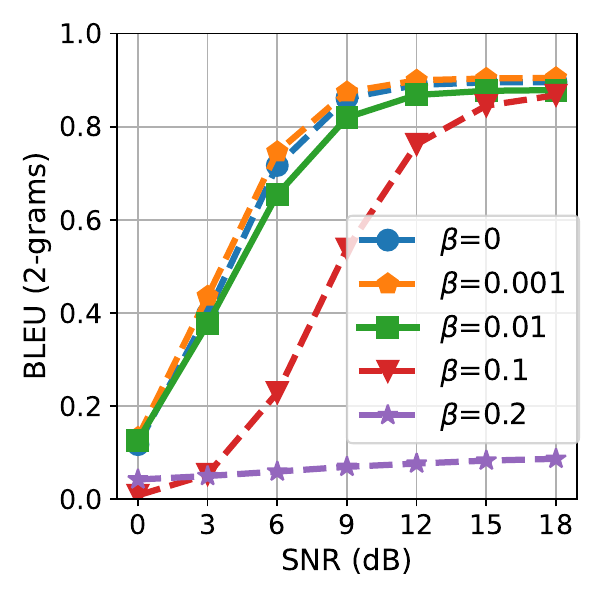}}
        \subfloat[Eve.]{
        \includegraphics[width=0.24\textwidth]{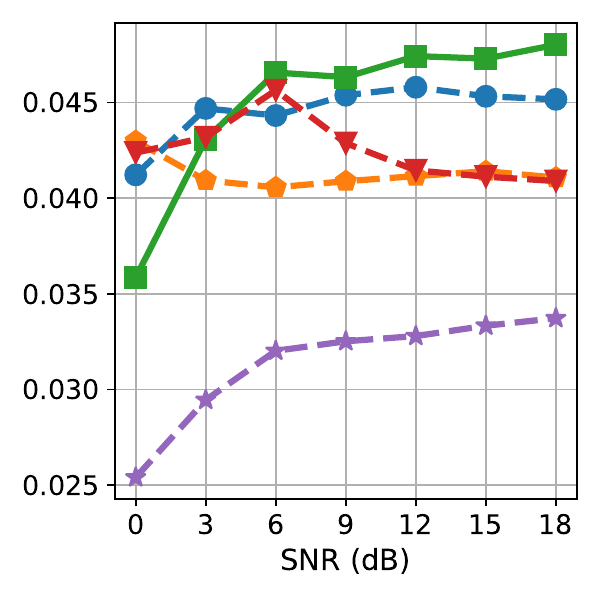}}
	\caption{\liyao{Bob's and Eve's BLEUs at $\alpha$=0.01 and varying $\beta$s.}}
	\label{fig:beta}
\end{figure}
\liyao{The results of fixing $\alpha=0.01$ and varying the value of $\beta$ are shown in Fig. ~\ref{fig:beta}, where the solid line represents results at the default value of $\beta=$0.01. Since $\beta$ controls the weight of $C_s$, it has significant impacts on both Bob's and Eve's performance. An overly large $\beta$ would result in less leakage at Eve, but also severely degrades Bob's performance.}

\liyao{Overall, in terms of absolute performance, increasing the values of $\alpha$ and $\beta$ both lead to a lower level of leakage for Eve, but negatively affect Bob's rates. Hence it is important to choose an appropriate parameter.
}

\textbf{Discussion.} 
Through the above analysis, we can conclude that encryption with shuffling is a kind of destruction of the original data distribution. If the shuffling is sufficiently random, little correlation would occur between the data before and after shuffling, almost achieving perfect secrecy. However, we also want the destruction to be reversible without any information loss. Taking advantage of the network structure, we can achieve theoretically reversible destruction, which is more explicit and explainable than the adversarial method. Thus our method ends up with a better semantic transmission performance with minimal leakage.

%% file: conclusion.tex
\section{Conclusion}
\label{sec:conslusion}
In this paper, we propose a secure semantic communication framework for the wiretap channel by novelly applying random shuffling to the intermediate representation of DNN as the shared secret key. The system aims at maximizing the transmission capacity between legitimate communication parties while suppressing the leakage to the eavesdropper. Experimental results indicate that our proposed method outperforms traditional methods in terms of robustness, transmission accuracy, and secrecy, on a variety of data transmission tasks.
%We proposed shuffling method is a hot-swappable module that can be flexibly adapted to existing neural network-based communication systems with little impact on accuracy and computational complexity. We joint design the loss function with security capacity and information leakage rate to improve the security and transmission efficiency of the system, which the latter is also calculated to evaluate the security performance of the system. 

%It is also suitable for various data reconstruction tasks and neural network structures. In the future, we will further explore the tradeoff between communication reliability and security. The encoding network of Alice is required to achieve two seemingly contradictory goals: a higher transmission rate and a lower information leakage rate. How to manipulate the encoding scheme (or the destruction of data distribution) to balance the two is an interesting direction to investigate in the future.

%% file: supplement.tex
\section*{Supplementary Material}
\subsection*{A. Proof of Eve's Channel Capacity}
\label{proofprocess}
\begin{IEEEproof}
    We define the information transmission rate $R$ as $R= \frac{log |\mathcal{M}|}{n}$, and hence
    \begin{equation}
    \label{ratedefinition}
    \begin{aligned}
         nR &= \log |\mathcal{M}| \\
          & \overset{(1)}{=} H(\boldsymbol{M}) \\
          & \overset{(2)}{=} H(\boldsymbol{M}|\boldsymbol{\hat{M}}) + I(\boldsymbol{M};\boldsymbol{\hat{M}}) \\
          & \overset{(3)}{\leq} 1 + P_{e}nR + I(\boldsymbol{U}^n ; \boldsymbol{X}^n), \\
    \end{aligned} 
    \end{equation}
    where (1) holds due to the assumption that $M$ is uniformly distributed, (2) follows the definition of mutual information, and (3) follows the \emph{Fano' inequality } and \emph{data processing inequality}. 
    
    Due to the shuffling operation, $I(\boldsymbol{U}^n; \boldsymbol{X}^n) \neq nI(\boldsymbol{U}; \boldsymbol{X})$, but we can directly calculate the mutual information between $\boldsymbol{U}^n$ and $\boldsymbol{X}^n$ by the definition of the shuffling operation.
    For $\boldsymbol{U}^n \to \boldsymbol{X}^n$, the transition probability matrix $P_{\boldsymbol{X}^n|\boldsymbol{U}^n}$ is:
    \begin{equation}
    \renewcommand{\arraystretch}{1.5} % 调整行间距
    \setlength{\arraycolsep}{2pt} % 调整列间距
    P_{\boldsymbol{X}^n|\boldsymbol{U}^n} = 
        \left[
            \begin{array}{cccc}
                p_{11} & p_{21} & \dots &p_{r1} \\
                p_{12} & p_{22} & \dots &p_{r2} \\
                \vdots &  \vdots&\ddots&\vdots \\
                p_{1m} & p_{2m} & \dots & p_{rm}                 
            \end{array}
        \right]
        =
        \left[
            \begin{array}{cccc}
                \frac{1}{a} & 0 &\dots&\frac{1}{a} \\
                \frac{1}{a} & 0&\dots&0 \\
                \vdots & \vdots&\ddots&\vdots \\
                0 & \frac{1}{a}&\dots&\frac{1}{a}                 
            \end{array}
        \right]
    \end{equation}
    where $m=\mathcal{|U|}^n$, $r=\mathcal{|X|}^n$, $\sum_{j=1}^{r} P(x_j^n|u_i^n)=1$, and $a=(\frac{n}{g})!$ denote the number of patterns that can be obtained by shuffling each $u_i^n$ of granularity $g$, so each row of $P_{\boldsymbol{X}^n|\boldsymbol{U}^n}$ has $a$ items are $\frac{1}{a}$ and the remaining $r-a$ items are 0, thus the matrix $P_{\boldsymbol{X}^n|\boldsymbol{U}^n}$ has $ma$ non-zero terms. 
    The grain $g$ means how many elements are bound as one basic shuffling unit, which is a factor of $n$, so $g, \frac{n}{g} \in \mathbb{N_+}$.
    The mutual information has a maximal value when the channel input is uniformly distributed, assuming $\forall i \in [1,m], P(u_i^n)=\frac{1}{m} = \frac{1}{\mathcal{|U|}^n} $, then we have 
    \begin{equation}
    \label{mutualinformation}
    \begin{aligned}
         I(\boldsymbol{U}^n;\boldsymbol{X}^n) &= H(\boldsymbol{X}^n) -H(\boldsymbol{X}^n|\boldsymbol{U}^n) \\
          %& \overset{(a)}{\leq}  \log{m} - H(S|X) \\
          %& \overset{(b)}{=} H(X) - H(X|S) \\
          & \overset{(a)}{\leq} \log{\mathcal{|X|}^n} + \sum_{i=1}^{m} \sum_{j=1}^{r} P(x_j^n|u_i^n)P(u_i^n)\log {P(x_j^n|u_i^n)} \\
          &= n\log{\mathcal{|X|}} + \frac{1}{m} \sum_{p_{ji}\ne 0} P(x_j^n|u_i^n)\log {P(x_j^n|u_i^n)} \\
          &= n\log{\mathcal{|X|}}  + \sum_{p_{ji}\ne 0} \frac{1}{ma}\log {\frac{1}{a}} \\
          &= n\log{\mathcal{|X|}}  - \log{a} \\
          &\overset{(b)}{\cong}  n\log{\mathcal{|X|}} -\log{\sqrt{\frac{2 \pi n}{g}}(\frac{n}{eg})^{\frac{n}{g}}},   \\
    \end{aligned}  
    \end{equation}
   where (a) follows the uniform-distribution assumption of $X^n$ and (b) follows Stirling's approximation on the factorial. Thus, substituting Eq.~\eqref{mutualinformation} into Eq.~\eqref{ratedefinition} completes the proof of Eq.~\eqref{rateeve}.
   
   The result above is obtained by shuffling one-dimensional sequences but could be extended to matrices or higher-order data by reshaping. For example, a sequence of length $n$ can be restructured into a matrix of shape $ k \times \frac{n}{k} $, where $k, \frac{n}{k} \in \mathbb{N}_{+}$, subsequently, shuffling operations can be applied across rows and columns of this matrix and thus $a = (\frac{n}{k})!k!$. It becomes evident that $a$ is a function of $n!$, expressed as $a = \Theta(n!)$. Consequently, while different permutation methods introduce variations in the $\log a$ term in Eq.~\eqref{mutualinformation}, they do not alter the conclusion that Eve's channel capacity is zero.

\end{IEEEproof}
\vspace{-0.4cm}
\subsection*{B. Network Architectures}
We show all the network architectures in Table~\ref{tab:networksetting}.
\begin{table}[h]
\renewcommand\arraystretch{1.3}
\caption{The network architecture setting.}
\setlength{\abovecaptionskip}{1.5pt}
\setlength{\tabcolsep}{1.3mm}
\centering
    \label{tab:networksetting}
    \begin{tabular}{|l|l|l|l|l|}
    \hline
     & Network & Layer Name & Units & Activation \\ \hline
    \multirow{7}{*}{Text} & \multirow{3}{*}{$T_{Alice}$} & Word Embedding & 128 & None \\ \cline{3-5} 
     &  & \begin{tabular}[c]{@{}l@{}}4$\times$Transformer\\ Encoder Layer\end{tabular} & 128 (8 heads) & Linear \\ \cline{3-5} 
     &  & 2$\times$Dense & 128-16 & Relu \\ \cline{2-5} 
     & \multirow{3}{*}{\begin{tabular}[c]{@{}l@{}}$T_{Bob}$ and \\ $T_{Eve}$\end{tabular}} & 2$\times$Dense & 16-128 & Relu \\ \cline{3-5} 
     &  & \begin{tabular}[c]{@{}l@{}}4$\times$Transformer\\ Decoder Layer\end{tabular} & 128 (8 heads) & Linear \\ \cline{3-5} 
     &  & Prediction Layer & Dictionary Size & Softmax \\ \cline{2-5} 
     & $T_{\theta}$ & 3$\times$Dense & 32-100-1 & Relu \\ \hline
    \multirow{6}{*}{Image} & \multirow{3}{*}{$T_{Alice}$} & Patch Embedding & 48 & None \\ \cline{3-5} 
     &  & \begin{tabular}[c]{@{}l@{}}6$\times$Transformer\\ Encoder Layer\end{tabular} & 48 (4 heads) & Linear \\ \cline{3-5} 
     &  & 2$\times$Dense & 48-16 & Relu \\ \cline{2-5} 
     & \multirow{2}{*}{\begin{tabular}[c]{@{}l@{}}$T_{Bob}$ and \\ $T_{Eve}$\end{tabular}} & 2$\times$Dense & 16-48 & Relu \\ \cline{3-5} 
     &  & \begin{tabular}[c]{@{}l@{}}6$\times$Transformer\\ Decoder Layer\end{tabular} & 48 (4 heads) & Linear \\ \cline{2-5} 
     & $T_{\theta}$ & 3$\times$Dense & 32-100-1 & Relu \\ \hline

    \multirow{6}{*}{Speech} & \multirow{3}{*}{$T_{Alice}$} & Input Layer & 1 & None \\ \cline{3-5} 
    &  & 6$\times$SE-ResNet & 128 & Relu \\ \cline{3-5} 
    &  & CNN Layer & 128 & None \\ \cline{2-5} 
    & \multirow{2}{*}{\begin{tabular}[c]{@{}l@{}}$T_{Bob}$ and \\ $T_{Eve}$\end{tabular}} & CNN Layer & 128 & Relu \\ \cline{3-5} 
    &  & 6$\times$SE-ResNet & 128-1 & Relu \\ \cline{2-5} 
    & $T_{\theta}$ & 3$\times$Dense & 2-10-1 & Relu \\ \hline
    \end{tabular}
\end{table}

\subsection*{\liyao{C. Process of Key Sharing}}
\label{sec:keyshared}

\begin{figure}[htb]
	\centering
	\includegraphics[width=\linewidth]{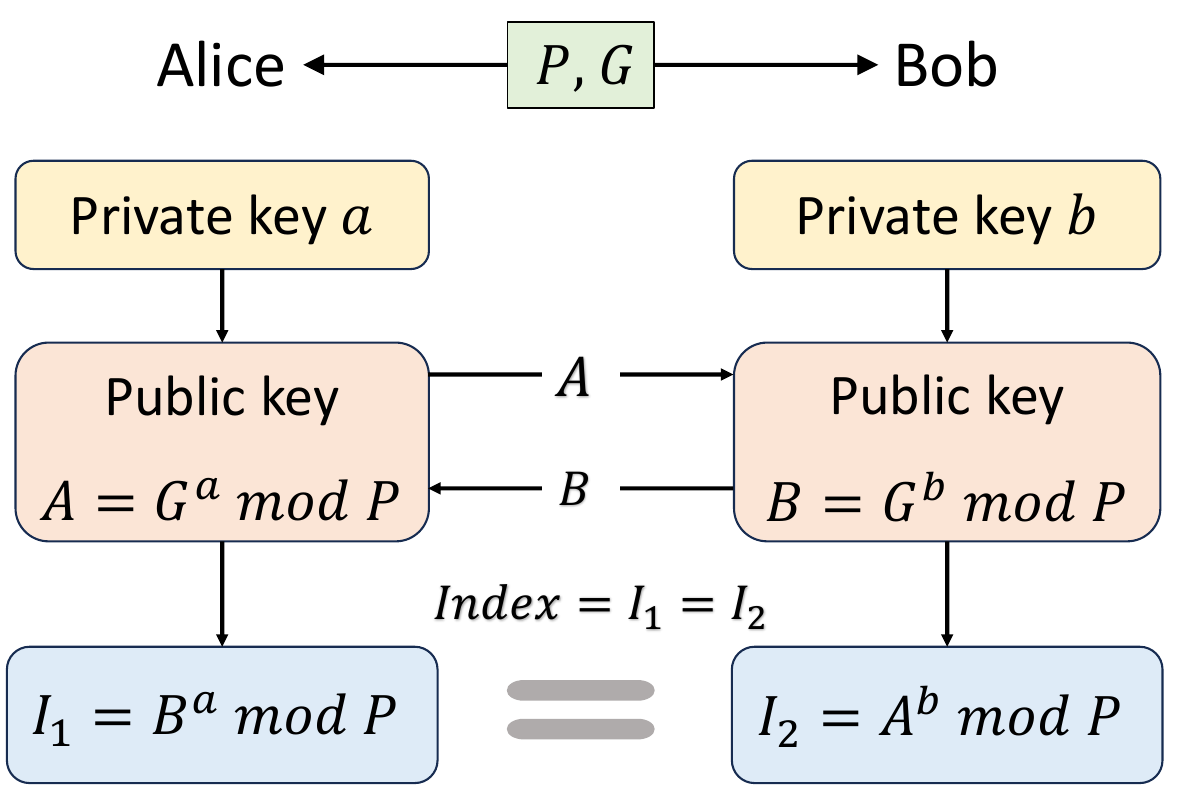}
	\caption{\liyao{Diffie-Hellman key exchange protocol. Alice and Bob share a codebook in prior and establish a common key by querying the codebook by the shared index.}}
	\label{dh}
\end{figure}

The Diffie-Hellman key exchange protocol can exchange keys securely over the public channel. The specific process of the key exchange protocol is shown in Fig. ~\ref{dh}.

\begin{enumerate}
	\item Alice and Bob negotiate on a large prime number $P$ and a primitive root $G$ of $P$, both of which are public global parameters.
	\item Alice generates the local private key $a$, computes the public key $A=G^{a}\, mod \,P$, and sends $A$ to Bob.
	\item Bob generates the local private key $b$, computes the public key $B=G^{b} \,mod \,P$, and sends $B$ to Alice.
	\item Alice receives the public key $B$ and calculates the key $I_{1}=B^{a}\,mod \,P$.
	\item Bob receives the public key $A$ and calculates the key $I_{2}=A^{b}\,mod \,P$
	\item Because of the  $(G^{a})^{b} \,mod\,P = (G^{b})^{a} \,mod\,P $, so Alice and Bob share the same secret number $Index=I_{1}=I_{2}$.
\end{enumerate}

Diffie-Hellman key exchange protocol is used to share a common index in the codebook. By selecting a large enough prime number $P$ in the protocol, the generated number shall cover the range of the index in the codebook. The shuffling key is retrieved according to the common index by both Alice and Bob.
\subsection*{D. Text Reconstruction Results under the Sentence Similarity Metric}
\label{sec:sentence_simi}
\begin{figure}[htbp]
	\centering
		\includegraphics[width=0.99\columnwidth]{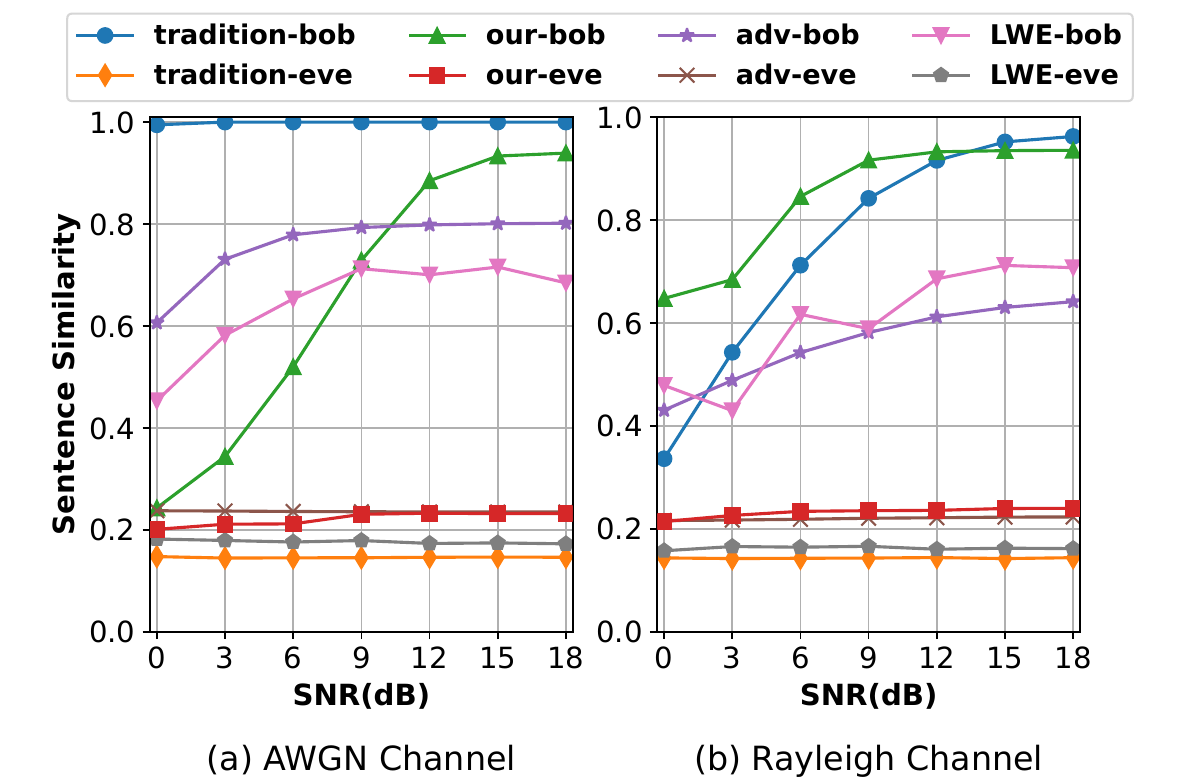}
	\caption{\liyao{Bob and Eve’s sentence similarity for different methods in (a) AWGN channel, and (b) Rayleigh fading channel.}}
	\label{sentence-similarity}
	% \vspace{-0.3cm}
\end{figure}
Fig.~\ref{sentence-similarity} displays the sentence similarity scores of four methods. The results mostly agree with that under BLEU. In particular, our method achieves a sentence similarity score over 0.8 at SNR greater than 6 dB in the Rayleigh channel, indicating a successful recovery of semantic information at the sentence level.

\subsection*{\lyr{E. Results on CIFAR-100 and CelebA-HQ}}

\lyr{In this section, we provide additional results on the CIFAR-100 and CelebA-HQ datasets.}

\label{sec:msssim}
\begin{figure}[htbp]
	\centering
	\subfloat[AWGN channel.]{
		\includegraphics[width=0.24\textwidth]{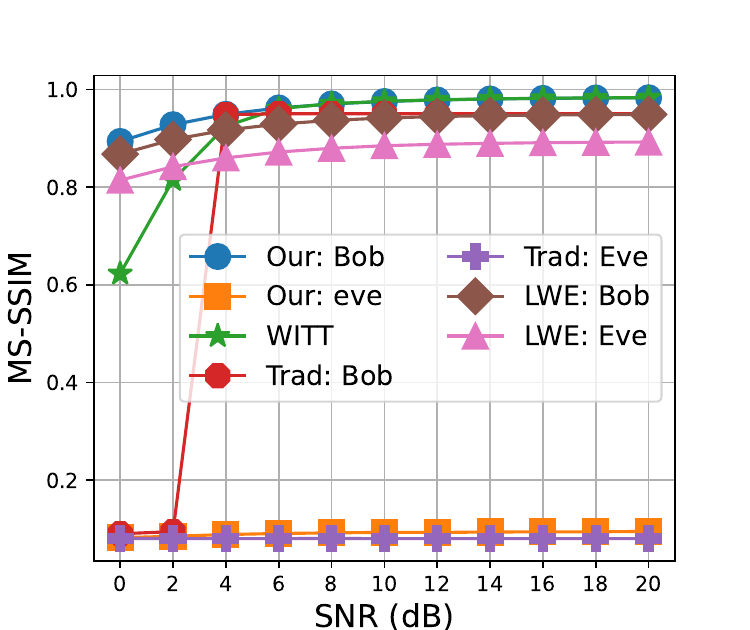}\label{MS-SSIM_AWGN}}
	\subfloat[Rayleigh fading channel.]{
		\includegraphics[width=0.24\textwidth]{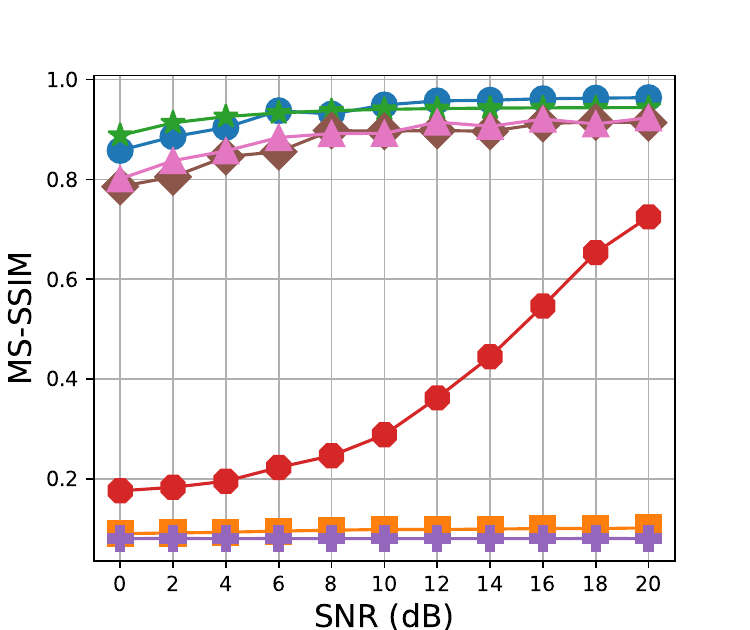}\label{MS-SSIM_Rayleigh}}
	\caption{\lyr{Image transmission on CelebA-HQ: Bob and Eve’s MS-SSIM Scores for different methods in (a) AWGN channel and (b) Rayleigh channel}.}
	\label{MSSIMscore}
\end{figure}

\lyr{Fig.~\ref{MSSIMscore} displays the MS-SSIM score on CelebA-HQ versus SNRs across different channels, where our method consistently demonstrates superior performance. The reconstruction performance of Bob in LWE is closer to our method and WITT in MS-SSIM than in PSNR, primarily because deep learning-based methods capture semantic information rather than bit errors, demonstrating the effectiveness of this technique.} Eve's MS-SSIM under our method, again, has a close-to-zero value indicating failure in reconstructing the images, while LWE remains broken under the attack.

\begin{figure}[h]
    \centering    \includegraphics[width=0.95\linewidth]{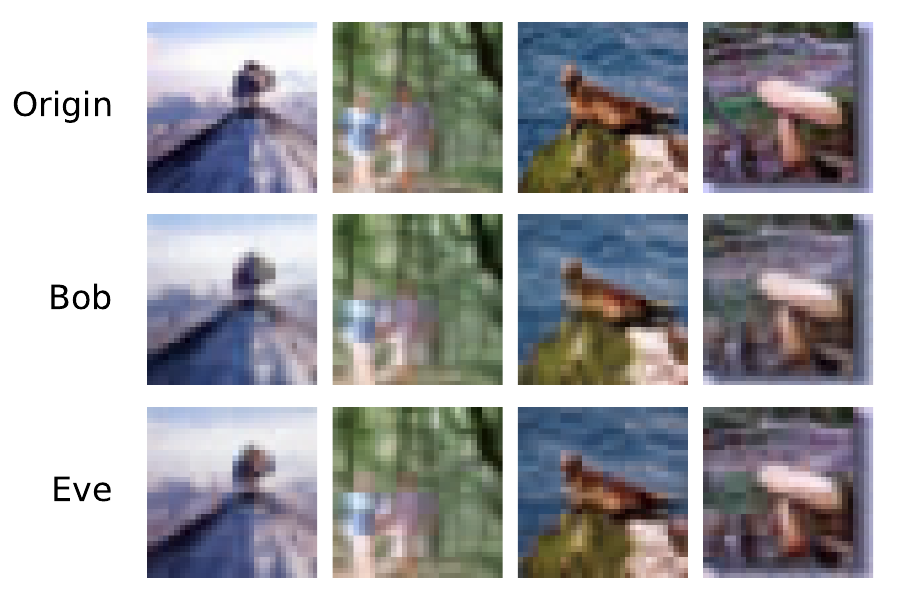}
    \caption{\liyao{Visualization of Eve's inversion attack results without any protection on CIFAR-100. } }
    \label{attack_visual_celeba}
\end{figure}

\lyr{Fig.~\ref{attack_visual_celeba} presents the inversion attack results on unprotected CIFAR-100, following a similar setting to that of Fig.~\ref{attack_visual}. The results confirm that eavesdropper Eve can successfully reconstruct original messages via inversion attacks in the absence of protection.}

% \begin{figure}[htbp]
%     \centering
%     \includegraphics[width = 0.85\linewidth]{Figure/keyspace_line.pdf}
%     \caption{\liyao{Image transmission on CIFAR-100: PSNR scores of Eve in different shuffling modes.}}
%     \label{keyspace_line_celeba}
% \end{figure}
% \begin{figure}[htbp]
%     \centering
%     \includegraphics[width = 0.99\linewidth]{Figure/keyspace.pdf}
%     \caption{\liyao{Image transmission on CIFAR-100: visualization of Eve's reconstruction in different shuffling modes.}}
%     \label{keyspace_celeba}
% \end{figure}

\begin{figure}[htbp]
    \centering
    \includegraphics[width = 0.85\linewidth]{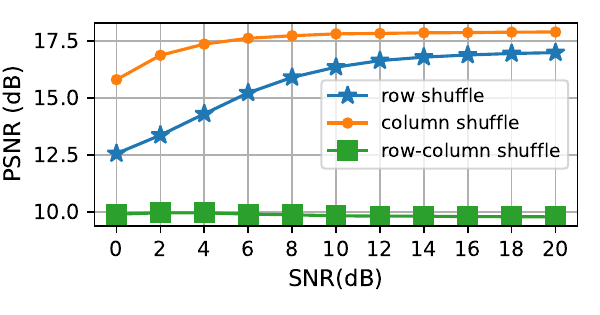}
    \caption{\lyr{Image transmission on CelebA-HQ: PSNR scores of Eve in different shuffling modes.}}
    \label{keyspace_line_celeba}
\end{figure}
\begin{figure}[htbp]
    \centering
    \includegraphics[width = 0.99\linewidth]{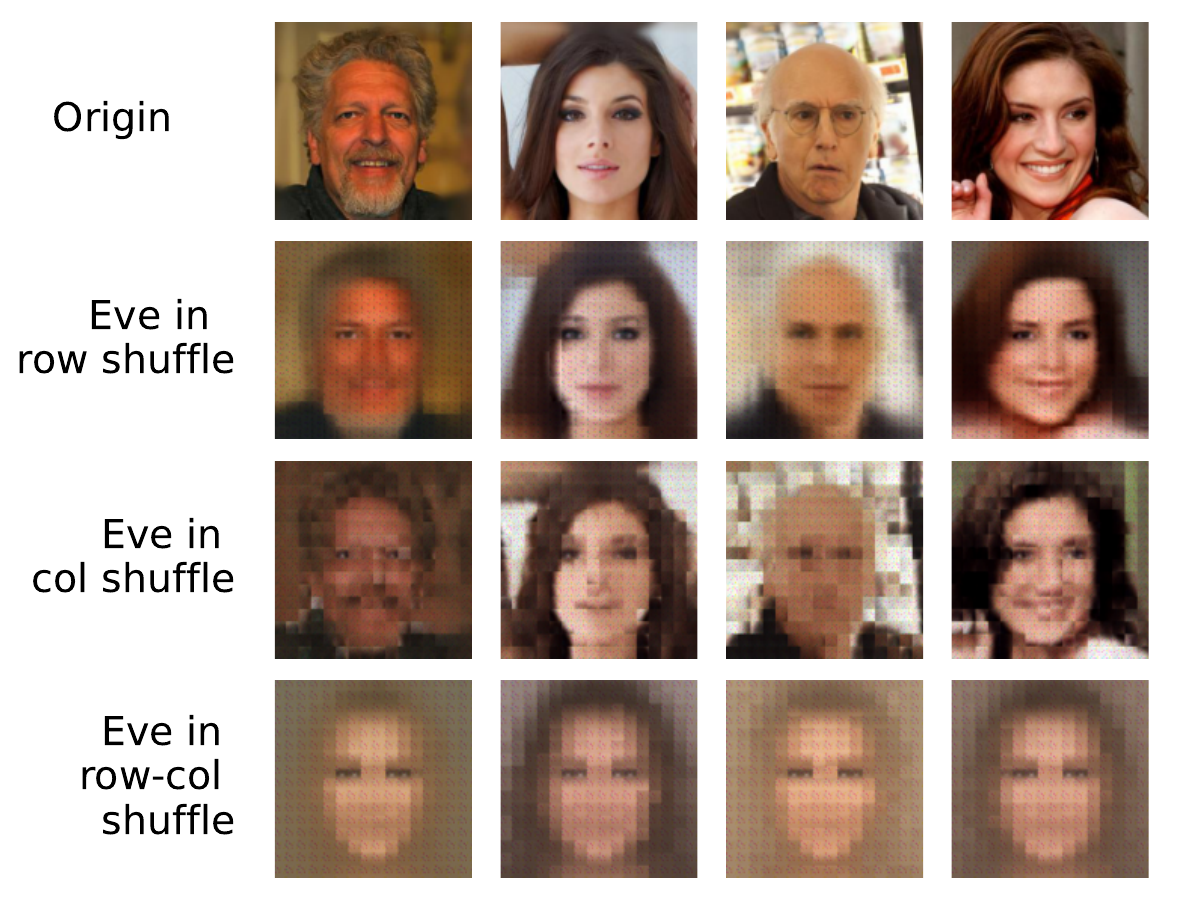}
    \caption{\lyr{Image transmission on CelebA-HQ: visualization of Eve's reconstruction in different shuffling modes.}}
    \label{keyspace_celeba}
\end{figure}

\lyr{Fig.~\ref{keyspace_line_celeba} shows Eve's PSNR scores in different shuffling modes, which are counterparts to those of Fig.~\ref {keyspace_line}. Fig.~\ref {keyspace_celeba} shows the visualization of Eve’s reconstruction in different shuffling modes on CelebA-HQ. It demonstrates that, under pure row shuffling, Eve can roughly recover the outline of the image with fine details missing (e.g., an elderly male face is reconstructed as a middle-aged male). This is because row shuffling disrupts inter-patch relationships while preserving some structural information. With column shuffling, Eve can reconstruct coarser but more recognizable images, as shuffling merely takes place within a patch, and thus the reconstructed image presents a mosaic-like effect. When both row and column shuffling are applied, Eve fails to recover any meaningful image but only produces an average face pattern, further confirming the security of the proposed scheme.}

% \lyr{Fig.~\ref{keyspace} shows that under pure row shuffling, Eve can roughly recover the outline of the image with fine details missing (e.g., an elderly male face is reconstructed as a middle-aged male). This is because row shuffling disrupts inter-patch relationships while preserving some structural information. With column shuffling, Eve can reconstruct coarser but more recognizable images, as shuffling merely takes place within a patch, and thus the reconstructed image presents a mosaic-like effect. When both row and column shuffling are applied, Eve fails to recover any meaningful image but only produces an average face pattern, further confirming the security of the proposed scheme.}

\begin{figure}[htbp]
	\centering
	\subfloat[AWGN channel.]{
		\includegraphics[width=0.24\textwidth]{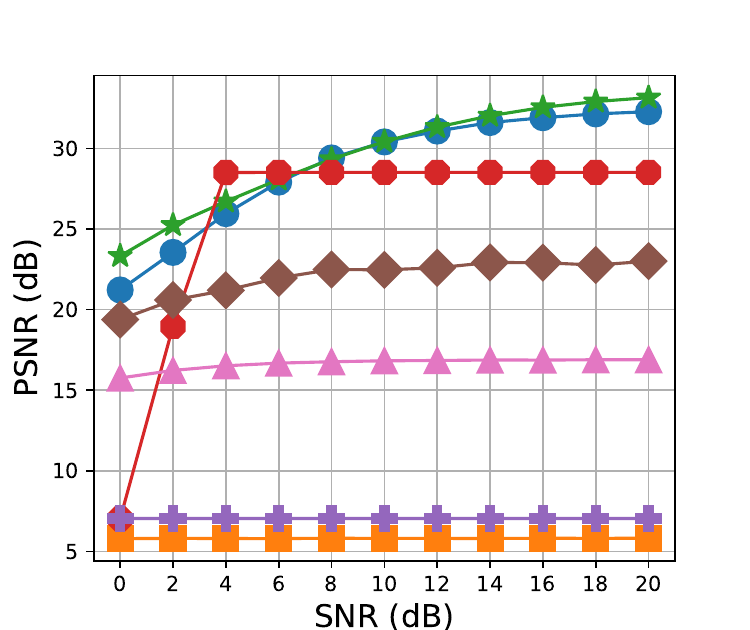}\label{PSNR_AWGN_celeba}}
	\subfloat[Rayleigh fading channel.]{
		\includegraphics[width=0.24\textwidth]{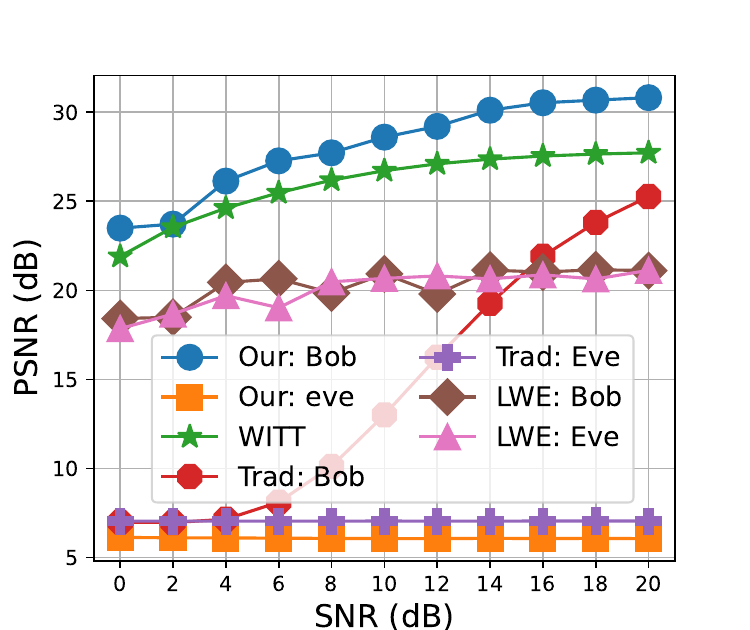}\label{PSNR_Rayleigh_celeba}}
	\caption{Image transmission \lyr{on CIFAR-100}: Bob and Eve’s PSNR scores for different methods in (a) AWGN channel and (b) Rayleigh channel.}
	\label{PSNR_celeba_celeba}
\end{figure}

\begin{figure}[htbp]
	\centering
	\subfloat[AWGN channel.]{
		\includegraphics[width=0.24\textwidth]{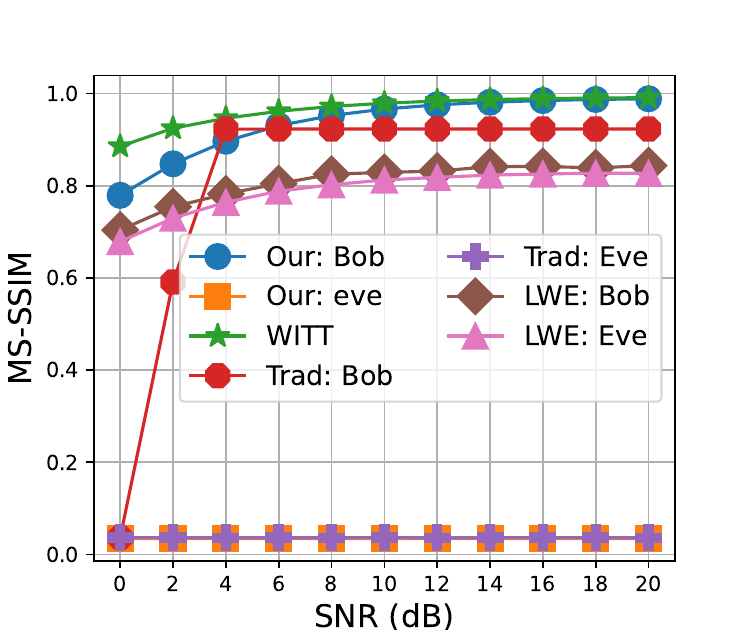}\label{MS-SSIM_AWGN_celeba}}
	\subfloat[Rayleigh fading channel.]{
		\includegraphics[width=0.24\textwidth]{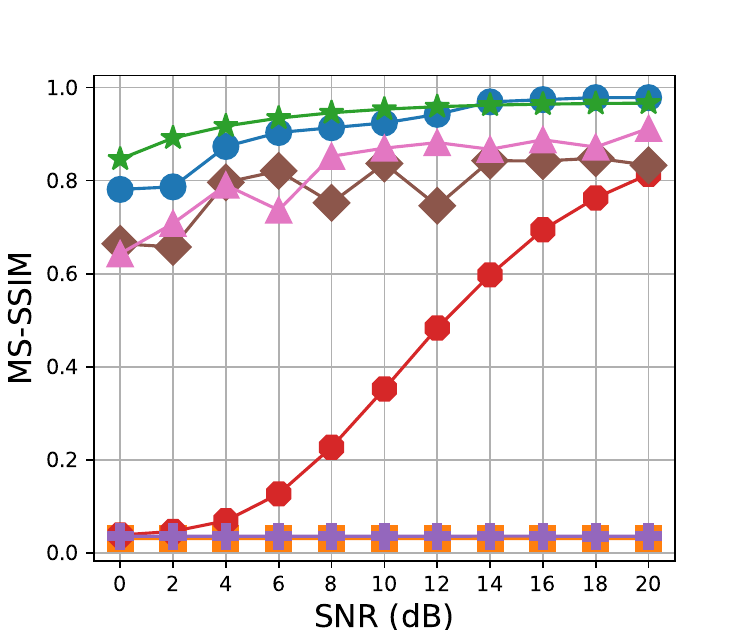}\label{MS-SSIM_Rayleigh_celeba}}
	\caption{Image transmission \lyr{on CIFAR-100}: Bob and Eve’s MS-SSIM Scores for different methods in (a) AWGN channel and (b) Rayleigh channel.}
	\label{MSSIMscore_celeba_celeba}
\end{figure}

\lyr{Fig.~\ref{PSNR_celeba_celeba} reports the PSNR scores of the CIFAR-100 dataset under AWGN and Rayleigh channels. The results for the AWGN channel demonstrate trends consistent with those in Fig.~\ref{fig:psnr}; only the numerical values are lower due to the use of lower-resolution images. For the Rayleigh channel results, deep learning-based methods exhibit better performance, while traditional methods still perform poorly. Fig.~\ref {MSSIMscore_celeba_celeba} displays the MS-SSIM score of the CIFAR-100 dataset versus SNRs across different channels. Different from PSNR, the MS-SSIM score of the traditional method under the AWGN channel is close to that of the semantic approach due to the low resolution of CIFAR-100 images. In general, the trend of the MS-SSIM score is similar to that under the PSNR metric.} 

\begin{figure}[htbp]
	\centering
	\subfloat[AWGN channel.]{
		\includegraphics[width=0.24\textwidth]{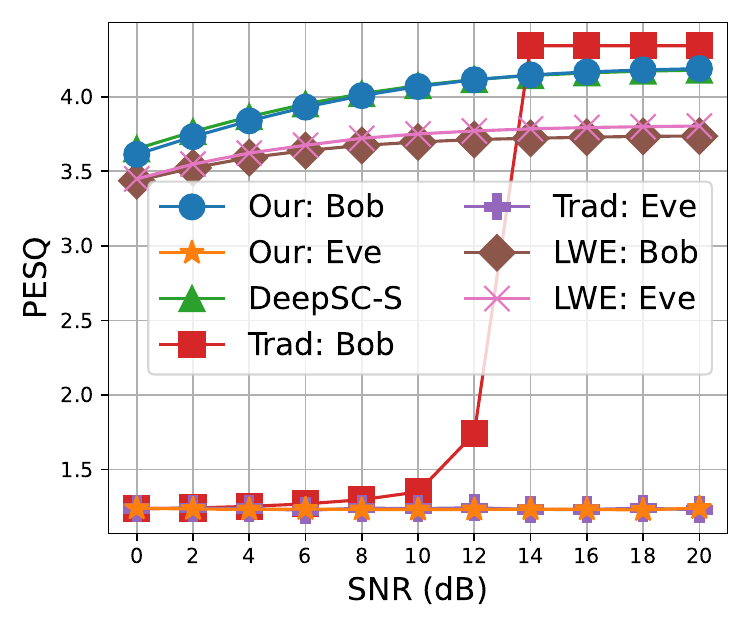}\label{PESQ_AWGN}}
    \subfloat[Rayleigh fading channel.]{
        \includegraphics[width=0.24\textwidth]{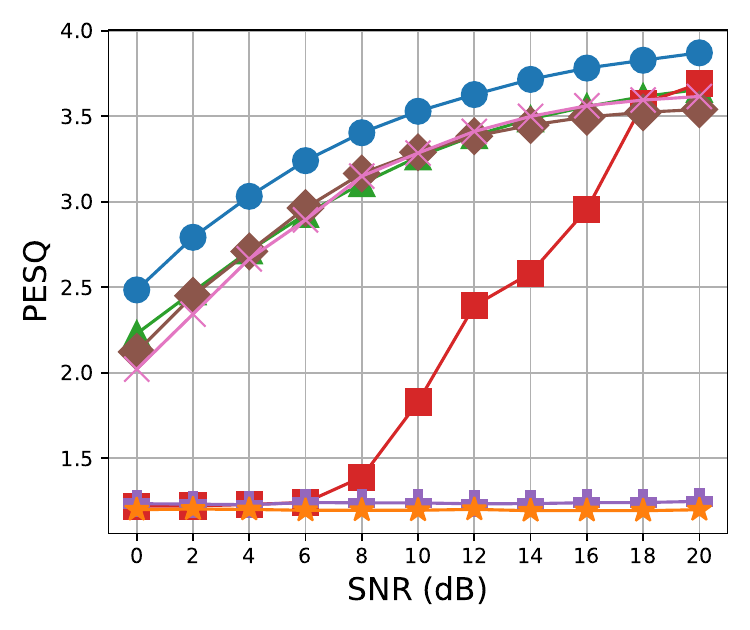}\label{PESQ_Rayleigh}}
	\caption{Speech transmission: Bob and Eve’s PESQ score for different methods in (a) AWGN channel and (b) Rayleigh channel.}
	\label{PESQscore}
\end{figure}

\subsection*{F. Speech Reconstruction Results under the PESQ Metric}
\label{sec:pesq}

% PESQ performance
Fig.~\ref{PESQscore} illustrates the PESQ score versus SNR across both channels, mirroring the trends observed in the SDR performance. While the traditional method achieves optimal results in channels with high SNRs, it is consistently inferior to our approach. The performance of traditional methods in Rayleigh channels is constantly worse than our method. Eve's results again prove that little leakage incurred with our approach, functioning in a similar way to the one-time pad in the traditional method. \liyao{LWE fails to maintain robustness against Eve's attack.}

\subsection*{G. Summary of Differences}
This work is an extension to our conference paper: Deep Learning Enabled Semantic-Secure Communication with Shuffling, published at the IEEE Global Communications Conference (GlobeCom) in the year 2023. In the below, we summarize the new contributions we made:
\begin{itemize}
    \item We extend the original text transmission system to a general one that incorporates images, texts, and audio transmissions (in Sec. I, III, IV, VI).
    \item We specifically introduce the key-sharing method to Sec. III-B.
    \item We provide a vivid example of permutation and inverse permutation in Fig. 3.
    \item We analyze the keyspace of our proposed approach in Sec. IV-A.
    \item We add a thorough leakage analysis in Sec. V to emphasize the security of the proposed framework. This leakage analysis is related to the capacity analysis of noisy permutation channels, which offers a novel view of our system.
    \item Extensive experiments on different types of data (e.g., images, audio speech), along with the corresponding baselines, are designed and added to Sec. VI. We also add ablation studies, including key rates, shuffling modes, weight factor, etc., to Sec. VI-G.
\end{itemize}